\documentclass[11pt]{article}
\usepackage{amssymb}
\usepackage{amsmath}
\usepackage[all]{xy}
\usepackage{graphicx}

\title{The Logos Categorical Approach to Quantum Mechanics: III.\\Relational Potential Coding and Quantum Entanglement\\Beyond Collapses, Pure States and Particle Metaphysics.}

\author{{\sc C. de Ronde}$^{1,2}$ and {\sc C. Massri}$^{3,4}$}
\date{}

\usepackage[margin=2.5cm]{geometry}

\begin{document}

\bibliographystyle{plain}
\maketitle

\begin{center}
\begin{small}
1. Philosophy Institute Dr. A. Korn (UBA-CONICET)\\
2. Center Leo Apostel for Interdisciplinary Studies\\Foundations of the Exact Sciences (Vrije Universiteit Brussel)\\
3. Institute of Mathematical Investigations Luis A. Santal\'o (UBA-CONICET)\\
4. University CAECE
\end{small}
\end{center}

\bigskip

\begin{abstract}
\noindent In this paper we consider the notion of {\it quantum entanglement} from the perspective of the logos categorical approach \cite{deRondeMassri18a, deRondeMassri18b}. Firstly, we will argue that the widespread distinctions, on the one hand, between {\it pure states} and {\it mixed states}, and on the other, between {\it separable states} and {\it entangled states}, are completely superfluous when considering the orthodox mathematical formalism of QM. We will then argue that the introduction of these distinctions within the theory of quanta is due to another two completely unjustified metaphysical presuppositions, namely, the idea that there is a ``collapse'' of quantum states when being measured and the idea that QM talks about ``elementary particles''. At distance from these distinctions and taking  the logos approach as a standpoint, we will propose an objective formal account of the notion of entanglement in terms of {\it potential coding} which introduces the necessary distinction between {\it intensive relations} and {\it effective relations}. We will also discuss how this new definition of entanglement provides also an {\it anschaulich} content to this ---supposedly ``spooky''--- quantum relational feature.
\end{abstract}
\begin{small}

\end{small}

\newtheorem{theo}{Theorem}[section]
\newtheorem{definition}[theo]{Definition}
\newtheorem{lem}[theo]{Lemma}
\newtheorem{met}[theo]{Method}
\newtheorem{prop}[theo]{Proposition}
\newtheorem{coro}[theo]{Corollary}
\newtheorem{exam}[theo]{Example}
\newtheorem{rema}[theo]{Remark}{\hspace*{4mm}}
\newtheorem{example}[theo]{Example}
\newcommand{\proof}{\noindent {\em Proof:\/}{\hspace*{4mm}}}
\newcommand{\qed}{\hfill$\Box$}
\newcommand{\ninv}{\mathord{\sim}} %involutive negation
\newtheorem{postulate}[theo]{Postulate}

\bigskip

\bigskip

\bigskip

\section*{Introduction}

The notion of {\it entanglement} plays today the most central role in what might be regarded as the origin of a new groundbreaking technological era. In the specialized literature, this research field falls under the big umbrella of what is called {\it quantum information processing}. This term covers different amazing, outstanding and completely non-classical technical developments such as, for example, quantum teleportation, quantum computation and quantum cryptography. Without exception, all these new technologies are founded on the notions of {\it quantum superposition} and {\it entanglement}. But even though technicians, computer scientists, cryptographers and engineers are rapidly advancing in deriving new technical devices and algorithms, the understanding of the physical concept of {\it entanglement} has remained almost completely untouched since its coming to life in 1935 when Albert Einstein, Boris Podolsky and Nathan Rosen discussed what would later become known as the famous EPR {\it Gedankenexperiment}. In that same year, Erwin Schr\"odinger in a series of papers gave its name to the new born concept. He discussed in depth what he called {\it entanglement} ({\it Verschr\"rankung} in German) and showed through the now famous `cat paradox' its non-classical features. The new notion was regarded with horror by the community of physicists. ``Spooky'' was the name they invented to bully the still very young notion. But after half a century, with the new technical possibilities, what Einstein, Podolsky, Rosen and Schr\"odinger had critically imagined, was now verifiable in the lab. The results, against their too classical physical intuitions, confirmed the predictions of the theory of quanta. Entanglement grow then rapidly, becoming one of the key concepts used within today's quantum technology. And as in the story of Hans Christian Andersen, the ugly duckling became the most powerful swan. 

In this article we attempt to provide an understanding of {\it quantum entanglement} through our new born logos categorical approach to QM presented in \cite{deRondeMassri18a, deRondeMassri18b}. It is organized as follows. In section 1 and 2, we address the very origin of the notion of quantum entanglement, firstly, implicitly presented within the EPR {\it Gedankenexperiment}, and secondly, explicitly addressed in a series of papers by Erwin Schr\"odinger. In section 3, we address the emergence and rapid growth in the 80's of quantum information processing from Aspect's famous experiment which finally tested the Boole-Bell inequality. Section 4 discusses representational realism as the philosophical standpoint of the logos approach. In section 5 we recall some basic notions of the logos categorical formalism. In section 6 we consider quantum entanglement from the perspective of the logos approach. Section 7 discusses how the logos scheme restores an objective account of entanglement in terms of {\it potential coding.} In section 8 we provide an intuitive ({\it aschaulich}) account of how to think about quantum entanglement. Finally, we provide some general conclusions.

\section{A Critical Analysis of EPR's 1935 {\it Gedankenexperiment}}

Critical thought is, above all, the possibility of analysis of the foundation of thought itself. The analysis of the conditions under which thinking becomes possible. By digging deeply into the basic components of thinking, one is able to understand the preconditions and presuppositions which support the architecture of argumentation itself. To provide a critical analysis of the EPR {\it Gedankenexperiment} presented in \cite{EPR} means to discuss the very foundations on which the argument presented by Einstein, Podolsky and Rosen grounds itself. Such type of analysis has been of course already provided within the foundational literature ---between many others--- by Diederik Aerts \cite{Aerts84a, Aerts84b} and Don Howard \cite{Howard85}. In the following, we attempt to extend this analysis paying special attention to the notion of separability, to the famous definition of {\it element of physical reality} and, of course, to their conclusion.   

\subsection{Einstein's (Spatio-Temporal) Separability Principle}

Even though Albert Einstein was certainly a revolutionary in many aspects of his research, he was also a classicist when considering the preconditions of physical theories themselves. His dream to create a unified field theory was grounded in his belief that physical theories, above all, must always discuss in terms of specific situations happening within space and time. In this respect, the influence of transcendental philosophy in Einstein's thought cannot be underestimated \cite{Howard94}. That space and time are the {\it forms of intuition} that allow us to discuss about objects of experience was one of the most basic {\it a priori} dictums of Kantian metaphysics, difficult to escape even for one of the main creators of relativity theory.

In a letter to Max Born dated 5 April, 1948, Einstein writes:
\begin{quotation}
\noindent {\small ``If one asks what, irrespective of quantum mechanics, is characteristic of the world of ideas of physics, one is first stuck by the following: the concepts of physics relate to a real outside world, that is, ideas are established relating to things such as bodies, fields, etc., which claim a `real existence' that is independent of the perceiving subject ---ideas which, on the other hand, have been brought into as secure a relationship as possible with the sense-data. It is further characteristic of these physical objects that they are thought of as arranged in a space-time continuum. An essential aspect of this arrangement of things in physics is that they lay claim, at a certain time, to an existence independent of one another, provided these objects `are situated in different parts of space'. Unless one makes this kind of assumption about the independence of the existence (the `being-thus') of objects which are far apart from one another in space ---which stems in the first place in everyday thinking--- physical thinking in the familiar sense would not be possible. It is also hard to see any way of formulating and testing the laws of physics unless one makes a clear distinction of this kind.'' \cite[p. 170]{Born71}}
\end{quotation}

\noindent This precondition regarding objects situated in different parts of space can be expressed, following Howard \cite[p. 226]{Howard89}, as a principle of spatio-temporal separability:

\smallskip
\smallskip

\noindent {\it {\bf Separability Principle:} The contents of any two regions of space separated by a non-vanishing spatio-temporal interval constitute separable physical systems, in the sense that (1) each possesses its own, distinct physical state, and (2) the joint state of the two systems is wholly determined by these separated states.}

\smallskip
\smallskip

\noindent In other words, the presence of a non-vanishing spatio-temporal interval is a sufficient condition for the individuation of physical systems and their associated states. Everything must ``live'' within space-time; and consequently, the characterization of every system should be discussed in terms of {\it yes-no questions} about physical properties. But, contrary to many, Einstein knew very well the difference between a conceptual presupposition of thought and the conditions implied by mathematical formalisms. In this respect, he also understood that his principle of separability was {\it only for him} a necessary metaphysical condition for doing physics. More importantly, he was aware of the fact there was no logical inconsistency in dropping the separability principle in the context of QM. At the end of the same letter to Born he points out the following:
\begin{quotation}
\noindent {\small ``There seems to me no doubt that those physicists who regard the descriptive methods of quantum mechanics as definite in principle would react to this line of thought in the following way: they would drop the requirement for the independent existence of the physical reality present in different parts of space; they would be justified in pointing out that the quantum theory nowhere makes explicit use of this requirement.'' \cite[p. 172]{Born71}}
\end{quotation}

\noindent This passage shows that Einstein was completely aware of the fact that QM is not necessarily committed to the metaphysical presupposition of space-time separability. But let us now turn to the kernel of the EPR argument, namely, their discussion regarding the sufficient conditions for defining what must be considered as physically real.

\subsection{EPR's {\it Sufficient Condition} for Physical Reality}

The notion of physical reality is of course the key element within the EPR line of argumentation and reasoning. Already in the first page of the paper, \cite{EPR}, Einstein, Podolsky and Rosen introduce the following ``reality criterion'' which stipulates a sufficient condition for considering an element of physical reality:\footnote{We are thankful to Prof. Don Howard for pointing us the specificity of the reality criterion.}

\smallskip
\smallskip

\noindent {\it {\bf Element of Physical Reality:} If, without in any way disturbing a system, we can predict with certainty (i.e., with probability equal to unity) the value of a physical quantity, then there exists an element of reality corresponding to that quantity.}

\smallskip
\smallskip

The relation drawn by the criterion is that between a {\it certain prediction} on the one hand, and the value of a physical quantity (or property) of the system on the other. Certainty is understood as ``probability equal to unity''. Notice that this remark is crucial in order to filter the predictions provided by QM. Only those related to probability equal to one, $p = 1$, can be considered to be related to physical reality. This means, implicitly, that the rest of the quantum mechanical probabilistic predictions which are not equal to one ---namely, those which pertain the interval between 0 and 1---, $p \in (0,1)$, are simply not considered. Given a quantum state, $\Psi$, there is only one meaningful operational statement (or property) that can be predicted with certainty. This leads to the conclusion that only one property can be regarded as being {\it actual}. The rest of quantum properties are considered as being {\it indeterminate}. The important point is that the ``non-certain'' predictions are not directly related to physical reality. Unlike real actual properties, indeterminate properties are considered as being only ``possible'' or ``potential'' properties; i.e., properties that might become actual in a future instant of time (see for a detailed analysis \cite{Sudbery16}). Until these properties are not actualized they remain in an unclearly defined limbo, in the words of Heisenberg \cite[p. 42]{Heis58}, they stand ``in the middle between the idea of an event and the actual event, a strange kind of physical reality just in the middle between possibility and reality.'' The filtering of indeterminate properties ---something which, from an operational perspective, seems completely unjustified---, is directly related to the actualist spatio-temporal (metaphysical) understanding of physical reality which Einstein so willingly wanted to retain. As he made the point \cite{Howard17}:  ``that which we conceive as existing (`actual') should somehow be localized in time and space. That is, the real in one part of space, $A$, should (in theory) somehow `exist' independently of that which is thought of as real in another part of space, $B$.'' As we discussed in detail in \cite{deRondeMassri18a}, it is not so difficult to see ---if we dig a bit deeper--- that this actualist understanding of existence is grounded in the classical representation of physics provided in terms of an {\it actual state of affairs} and {\it binary valuations}. 

But, as noticed by Bohr himself in his famous reply to EPR \cite{Bohr35}, it is the first part of the definition which introduces a serious ``ambiguity''. Indeed, the previous specification, {\it ``If, without in any way disturbing a system,''} refers explicitly to the possibility of measuring the system in question. It thus involves an improper scrambling between ontology and epistemology, between physical reality and measurement. A scrambling ---let us stress---, completely foreign to all classical physics. This scrambling, might be regarded as one between the many ``quantum omelettes''\footnote{The term ``quantum omelette'' which has become relatively famous in the foundational literature is due to Jaynes \cite[p. 381]{Jaynes} who argued in an uncontroversial manner: ``[O]ur present [quantum mechanical] formalism is not purely epistemological; it is a peculiar mixture describing in part realities of Nature, in part incomplete human information about Nature ---all scrambled up by Heisenberg and Bohr into an omelette that nobody has seen how to unscramble. Yet we think that the unscrambling is a prerequisite for any further advance in basic physical theory. For, if we cannot separate the subjective and objective aspects of the formalism, we cannot know what we are talking about; it is just that simple.'' As we shall discuss this term might be also related to EPR's operational definition of an {\it element of physical reality.}} created during the early debates of the founding fathers. In fact, this criteria contradicts one of the most interesting characterizations of physical theories provided by Einstein himself. Indeed, according to Einstein \cite[p. 175]{Dieks88a} : ``[...] it is the purpose of theoretical physics to achieve understanding of physical reality which exists independently of the observer, and for which the distinction between `direct observable' and `not directly observable' has no ontological significance''. This is of course, even though ``the only decisive factor for the question whether or not to accept a particular physical theory is its empirical success.'' The physical representation of a physical theory is always {\it prior} to the possibility of epistemic inquiry of which `measurement' is obviously one of its main ingredients. Recalling Einstein's famous remark to Heisenberg \cite[p. 63]{Heis71}: ``It is only the theory which decides what can be observed.''

\subsection{Collapses and Spooky Actions at a Distance}

Today, classical texts that describe QM axiomatically begin stating that the mathematical interpretation of a quantum system is a Hilbert space, that pure states are represented by rays in this space, physical magnitudes by self-adjoint operators on the state space and that the evolution of the system is ruled by the Schr\"{o}dinger equation. Possible results of a given magnitude are the eigenvalues of the corresponding operator obtained with probabilities given by the Born rule. In general the state is mathematically represented as a linear superposition of eigenstates corresponding to different eigenvalues of the measured observables.  Since we never ``observe'' superpositions it is argued that the theory requires an extra postulate which provides the missing link between quantum superpositions and measurement outcomes. This requirement was explicitly considered by John von Neumann and Paul Dirac in their famous books at the beginning of the thirties. As von Neumann's \cite[p. 214]{VN} made the point: ``Therefore, if the system is initially found in a state in which the values of $\mathcal{R}$ cannot be predicted with certainty, then this state is transformed by a measurement $M$ of $\mathcal{R}$ into another state: namely, into one in which the value of $\mathcal{R}$ is uniquely determined. Moreover, the new state, in which $M$ places the system, depends not only on the arrangement of $M$, but also on the result of $M$ (which could not be predicted causally in the original state) ---because the value of $\mathcal{R}$ in the new state must actually be equal to this $M$-result''. Or in Dirac's words: ``When we measure a real dynamical variable $\xi$, the disturbance involved in the act of measurement causes a jump in the state of the dynamical system. From physical continuity, if we make a second measurement of the same dynamical variable $\xi$ immediately after the first, the result of the second measurement must be the same as that of the first'' \cite[p. 36]{Dirac74}.

The introduction of the collapse within the axiomatic formulation of the theory has produced a series of paradoxes of which entanglement has not been the exception. The EPR {\it Gedeankenexperiment} exposes this paradox in what seems to be a non-local influence when measuring one of the particles on the other distant entangled partner. Indeed, if one accepts the orthodox interpretation of QM according to which the measurement of a quantum superposition induces a ``collapse'' to only one of its terms, Einstein, Podolsky and Rosen then show that there then seems to exist a super-luminous transfer of information from one particle to the other distant partner. Einstein was of course clearly mortified by this seemingly non-local ``quantum effect'' which he called {\it ``spukhafte Fernwirkung''}, translated later as ``spooky action at a distance''. Once the entangled particles are separated, all their properties still remain {\it indeterminate}. But, the moment we perform a measurement of an observable in one of the particles we also find out instantaneously what is the value of the distant partner ---in case we would choose to measure the same observable. Thus, the ``collapse'' of one of the particles also produces the ``collapse'' of the other distant entangled particle. Every time we measure an observable in one of the particles, the other particle ---as predicted by QM--- will be found to possess a strictly correlated value.\footnote{Let us remark that observability is used in this case a sufficient condition to define reality itself. There is involved here a two sided definition of what accounts for physical reality, either in terms of computing the certainty of an outcome (= real) or by observing an outcome which was uncertain but became actual (= real).} Einstein was not only against this ``spooky action'', he was also against the addition of a subjectively produced ``collapse''. In this respect, Einstein is quoted by Everett \cite[p. 7]{OsnaghiFreitasFreire09} to have said that he ``could not believe that  a mouse could bring about drastic changes in the universe simply by looking at it''. Schr\"odinger would also criticize the addition of the collapse to the theory:

\begin{quotation}
\noindent {\small``But jokes apart, I shall not waste the time by tritely ridiculing the attitude that the state-vector (or wave function) undergoes an abrupt change, when `I' choose to inspect a registering tape. (Another person does not inspect it, hence for him no change occurs.) The orthodox school wards off such insulting smiles by calling us to order: would we at last take notice of the fact that according to them the wave function does not indicate the state of the physical object but its relation to the subject; this relation depends on the knowledge the subject has acquired, which may differ for different subjects, and so must the wave function.'' \cite[p. 9]{OsnaghiFreitasFreire09}} \end{quotation}

%To sum up, the conclusion is that the ``collapse'' of the state of one of the particles produces an instant ``super-luminous collapse'' in the state of the other distant entangled partner. This influence in the real existence of the distant particle goes obviously against the locality condition imposed by Einstein's separability principle. According to this principle, the measurements we perform in one system should not influence the reality of a separated distant system.  

\subsection{EPR's Conclusion: Incompleteness and Hidden Variables}

As it is well known, the conclusion of the EPR paper is that QM is an incomplete theory; it does not consider all the {\it elements of physical reality} the theory should talk about. In a nut shell, the line of reasoning in order to end up with such a fatal veredict for quantum theory runs as follows: 

\begin{enumerate}
\item[I.] A {\it complete theory} is one which takes into account all its {\it elements of physical reality}. 

\item[II.]  QM presents a limit, due to to Heisenberg's uncertainty relations, to the knowledge of complementary properties (of the same quantum system).

\item[III.] If one accepts as a {\it sufficient condition} the proposed definition of an {\it element of physical reality}, then one can argue that incompatible properties of the same quantum system must be regarded as being all {\it elements of physical reality}, simultaneously. 

\item[IV.] If incompatible complementary properties are elements of physical reality, this means they possess a definite value previous to measurement, thus contradicting Heisenberg's relations ---a cornerstone of orthodox QM itself.

\item[V.] Since QM is incapable of considering all the {\it elements of physical reality} (i.e., the definite valued properties) of the separated systems under study simultaneously (due to Heisenberg's relations), the theory is incomplete. 
\end{enumerate}

Einstein, Podolsky and Rosen conclude that QM is incomplete. However, they also maintain that it should be possible to ``complete the theory''. A new theory that considers ---unlike quantum theory--- all elements of physical reality present within an EPR type experiment could be developed in the future. It is with this hopeful wish that they end the paper.  
\begin{quotation}
\noindent {\small ``While we have thus shown that the wave function does not provide a complete description of the physical reality, we left open the question of whether or not such a description exists. We believe, however, that such a theory is possible.'' \cite[p. 555]{Schr35b}}
\end{quotation}

\noindent Thus, according to EPR, a description in terms of separable systems with definite values should be, in principle, possible to develop. This has been understood in the literature ---due to Bell's reading \cite{Bell64}--- as implying the existence of a ``hidden variable theory'' which is able to account for an {\it actual state of affairs} and restore in this way an actualist description of nature.

\section{Schr\"odinger's {\it Entanglement} of Systems (and Knowledge)}

As it is well known, the notion of {\it entanglement} was introduced by Erwin Schr\"odinger in a series of three papers during the years of 1935 and 1936 \cite{Schr35a, Schr35b, Schr36}. He used these articles to continue his reflections regarding the EPR example. For many years these ideas rested ---apparently--- unnoticed, silenced by the rest of the (Bohrian inclined) physics community like a shameful sin ---of the most powerful theory of them all--- that everybody knew existed but no one wanted to admit. In the first paper of the series, {\it Discussion of Probability Relations Between Separated Systems}, he begins right from the start by defining the physical meaning of {\it entanglement} in terms of interacting systems.  
\begin{quotation}
\noindent {\small ``When two systems, of which we know the states by their respective representatives, enter into temporary physical interaction due to known forces between
them, and when after a time of mutual influence the systems separate again, then
they can no longer be described in the same way as before, viz. by endowing each
of them with a representative of its own. I would not call that one but rather the
characteristic trait of quantum mechanics, the one that enforces its entire
departure from classical lines of thought. By the interaction the two representatives
(or $\psi$-functions) have become entangled.'' \cite[p. 555]{Schr35b}}
\end{quotation}

\noindent Immediately after, Schr\"odinger continues to explain how to {\it disentangle} the two systems through measurement: 
\begin{quotation}
\noindent {\small ``To disentangle them we must gather further information by experiment, although we knew as much as anybody could possibly know about all that happened. Of either system, taken separately, all previous knowledge may be entirely lost, leaving us but one privilege: to restrict the experiments to one only of the two systems. After reestablishing one representative by observation, the other one can be inferred simultaneously. In what follows the whole of this procedure will be called the {\it disentanglement}. Its sinister importance is due to its being involved in every measuring process and therefore forming the basis of the quantum theory of measurement, threatening us thereby with at least a {\it regressus in infinitum}, since it will be noticed that the procedure itself involves measurement.''\cite[p. 555]{Schr35b}}
\end{quotation}

\noindent Schr\"odinger is of course making reference here to the famous ``spooky action at a distance'',  according to which the measurement of one physical quantity in one of the systems seems to influence the definite value of the same physical quantity in the other distant system. Making explicit reference to the EPR paper, Schr\"odinger remarks that: 
\begin{quotation}
\noindent {\small ``Attention has recently \cite{EPR} been called to the obvious but very disconcerting fact that even though we restrict the disentangling measurements to one system, the representative obtained for the other system is by no means independent of the particular choice of observations which we select for that purpose and which by the way are entirely arbitrary. It is rather discomforting that the theory should allow a system to be steered or piloted into one or the other type of state at the experimenter's mercy in spite of his having no access to it.''\cite[p. 555-556]{Schr35b}}
\end{quotation}

Following the reality criteria proposed in the EPR paper, Schr\"odinger also assumed ---implicitly--- that {\it maximal knowledge} had to be understood as {\it certain knowledge}; i.e., as restricted to knowledge that involves probability equal to unity. As he remarks, the astonishing aspect of QM is that when two systems get entangled through a known interaction, the knowledge we have of the parts might anyhow decrease. This of course is something completely untenable in classical physics. 
\begin{quotation}
\noindent {\small ``If two separated bodies, each by itself known maximally, enter a situation in which they influence each other, and separate again, then there occurs regularly that which I have just called {\it entanglement} of our knowledge of the two bodies. The combined expectation-catalog consists initially of a logical sum of the individual catalogs; during the process it develops causally in accord with known law (there is no question of measurement here). The knowledge remains maximal, but at the end, if the two bodies have again separated, it is not again split into a logical sum of knowledges about the individual bodies. What still remains {\it of that} may have become less than maximal, even very strongly so.---One notes the great difference over against the classical model theory, where of course from known initial states and with known interaction the individual states would be exactly known."\cite[p. 161]{Schr35a}}
\end{quotation}

\noindent We shall return to this important reflections from the perspective of the logos approach. But before doing so, let us now jump to the rise of entanglement during the last decades as ``a source'' for developing quantum information processing. 

\section{Boole-Bell Inequalities, Aspect Experiments and the Rise of Quantum Information Processing}

As remarked by Jeffrey Bub \cite{Bub17}, ``[...] it was not until the 1980s that physicists, computer scientists, and cryptographers began to regard the non-local correlations of entangled quantum states as a new kind of non-classical resource that could be exploited, rather than an embarrassment to be explained away.''  The reason behind this shift in attitude towards {\it entanglement} is an interesting one. As Bub continues to explain: ``Most physicists attributed the puzzling features of entangled quantum states to Einstein's inappropriate `detached observer' view of physical theory, and regarded Bohr's reply to the EPR argument (Bohr, 1935) as vindicating the Copenhagen interpretation. This was unfortunate, because the study of entanglement was ignored for thirty years until John Bell's reconsideration of the EPR argument (Bell, 1964).''

After the triumph of Bohr in the ``EPR battle'' \cite{Bohr35, EPR}, the notion of entanglement was almost completely erased by the orthodox community of physicists now following the Copenhagen spell. This was until an Irish researcher called John Stewart Bell working at the Conseil Europ\'een pour la Recherche Nucl\'eaire (CERN), wrote in 1964 a paper entitled {\it On the Einstein-Podolsky-Rosen Paradox}. In this paper he was able to derive a set of statistical inequalities that restricted the correlations described by any classical local-realistic theory \cite{Bell64}. Let us remark an important point, the Bell inequality is a statistical statement which is based on a set of (metaphysical) presuppositions derived from our classical representation of the world, it is a statement about classical probability theory. Contrary to a widespread confusion in some research circles, Bell type inequalities do not talk about QM. This fact becomes evident once we learn that a mathematically equivalent inequality was already derived by George Boole in 1864, exactly one century before the now famous version by Bell, and ---obviously--- long before the birth of the theory of quanta itself. As Itamar Pitowsky made the point:
\begin{quotation}
\noindent {\small ``In the mid-nineteenth century George Boole formulated his `conditions of possible experience'. These are equations and inequalities that the relative frequencies of (logically connected) events must satisfy. Some of Boole's conditions have been rediscovered in more recent years by physicists, including Bell inequalities, Clauser Horne inequalities, and many others.'' \cite[p. 95]{Pitowsky94}}\end{quotation}

\noindent As remarked by Pitowsky, Boole's research problem can be rephrased in modern terminology as follows: we are given a set of rational numbers $p_{1}, p_{2}, ...p_{n}$ which represent the relative
frequencies of $n$ logically connected events. The problem is to specify necessary and sufficient conditions that these numbers can be realized as probabilities in {\it some} probability space. These
conditions were called by Boole, for obvious reasons, ``conditions of possible experience''. But a more accurate term today would be: {\it conditions of possible classical experience}.

Continuing with our story, the true breaking point for the recognition of quantum entanglement and the possibilities it implied for quantum information processing was the unwanted  result of the famous experiment performed in Orsay at the very beginning of the 1980s by Alain Aspect, Philippe Grangier and Gerard Roger \cite{AGR81, AGR82}. The result was that the Boole-Bell inequality was violated by pairs of entangled spin ``particles''. As a consequence,  against Einstein and Bell's physical intuition, the possibility for classical theories to account for such experience was completely ruled out. The experiment designed by Aspect and his team ---repeated countless times up to the present \cite{Bernien13, Hensen15}--- could not be described by any classical local-(binary) realistic\footnote{Even though the original term is ``realistic'', we prefer to add ``binary'' for reasons that will become evident in the forgoing part of the paper.} theory. The experiment was also a sign that entanglement had to be taken seriously. It was only then that quantum computation, quantum cryptography and quantum teleportation, were developed by taking {\it entanglement} as a resource \cite{Bub17}. The new notion began to rapidly populate the journals, labs and research institutions all around the world. The technological era of quantum information processing had woken up from its almost half century hibernation. An hibernation, let us not forget, mainly due to the uncritical attitude of the majority of physicists who believed that Bohr had already solved everything ---and there was no reason to engage in metaphysical questions regarding physical reality. 

In the following section we will address all we have just analyzed and discussed from a new realist viewpoint, representational realism, which has allowed us to develop the logos categorical approach to QM.

\section{Revisiting EPR's Conclusion and Schr\"odinger's Analysis from a Representational Realist Viewpoint}

The notion of {\it representation} is key when discussing about the difficult and problematic relationship between physical theories on the one hand, and {\it physis} (or reality) on the other. In his paper published in {\it Naturwissenschaften} in 1935, Schr\"odinger begins his analysis by providing a deep discussion of the meaning of `representation' within physics and its relation to what he calls ``natural objects''. 
\begin{quotation}
\noindent {\small ``Of natural objects, whose observed behavior one might treat, one sets up a representation ---based on the experimental data in one's possession but without handcuffing the intuitive imagination--- that is worked out in all details exactly, {\it much} more exactly that any experience, considering its limited extent, can ever authenticate. The representation is its absolute determinacy resembles a mathematical concept or a geometric figure which can be completely calculated from a number of {\it determining parts}. [...] Yet the representation differs intrinsically from a geometric figure in this important respect, that also in {\it time} as fourth dimension it is just as sharply determined as the figure in the three space dimensions." \cite[p. 152]{Schr35a}}
\end{quotation}

\noindent Quite soon in the text, he continues to take distance from naive realism, arguing that: ``Of course one must not think so literally, that in this way [i.e., by creating representations] one learns how things go in the real world. To show this one does not think this, one calls the precise thinking aid that one has created, an {\it image} or a {\it model}.'' Some pages later, he finally assumes an empiricist stance which grounds the rest of his analysis in assuming ---like logical-positivists--- observability as a {\it given} of experience: 
\begin{quotation}
\noindent {\small ``Reality resists imitation trough a model. So one lets go of naive realism and leans directly on the indubitable proposition that {\it actually} (for the physicist) after all is said and done there is only observation, measurement. Then all our physical thinking thenceforth has the sole basis and as sole object the results of measurements which can in principle be carried out, for we must now explicitly {\it not} relate our thinking any longer to any kind of reality or to a model. All numbers arising in our physical calculations must be interpreted as measurement results. But since we didn't just now come into the world and start to build up our science from scratch, but rather have in use a quite definite scheme of calculation, from which in view of the great progress in Q.M. we would less than ever want to be parted, we see ourselves forced to dictate from writing-table which measurements are in principle possible, that is, must be possible in order to support adequately our reckoning system.'' \cite[p. 156]{Schr35a}}
\end{quotation}

\noindent These remarks are deeply important in order to analyze and discuss the representational realist viewpoint, already presented in \cite{deRonde16b, deRonde17a}, which provides a different characterization ---to that of Schr\"odinger--- of physics and the role played by {\it representation} within theories. Representational realism attempts to capture some of the core ideas present partly ---in different degrees--- within Einstein's, Heisenberg's and Pauli's understanding of the content and reference of physical theories. 

The kernel of representational realism is that physical theories represent reality through the creation of interrelated mathematical-conceptual schemes which allow us to provide  a quantitive and qualitative meaning and understanding to the unity of experience. According to representational realism, physical representation comes before any other consideration regarding the very possibility of physical analysis. In line with Einstein, Heisenberg and Pauli, and contrary to Schr\"odinger's understanding, our standpoint is that there are no ``natural objects'' to be found in nature. There is no {\it passive referent} waiting to be apprehended, no {\bf x} to be ``captured'' by a model. Or in other words, there is nothing ``natural'' about the notion of `object'. The idea of a {\it sameness} is already part of metaphysics. Our realist standpoint is that the notion of `physical object' is constituted both formally and conceptually in terms of a particular representation. In fact, the notion of `object' might be regarded as one of the most important creations of thought. A {\it moment of unity} which through counterfactual reasoning and discourse allows us to make sense of multiple physical phenomena. The notion of `object' is in itself a conceptual creation, it is not a self-evident ``natural existent'' of reality which is waiting to be observed. Since there is no object {\it before} providing the conditions of objective representation, science can only begin its analysis of experience when considering a specific theoretical representation. Any scientific discourse must presuppose a representation of what is meant by a `state of affairs'. This is not ---at least for the realist--- something ``self evident'' but the very precondition for understanding phenomena. In this respect we might also recall Einstein's remark to Heisenberg that: ``It is only the theory which decides what can be observed.'' This, in fact, is ---as remarked by Einstein--- the really significant  philosophical achievement of Kant:  
\begin{quotation}
\noindent {\small``From Hume Kant had learned that there are concepts (as, for example, that of causal connection), which play a dominating role in our thinking, and which, nevertheless, can not be deduced by means of a logical process from the empirically given (a fact which several empiricists recognize, it is true, but seem always again to forget). What justifies the use of such concepts? Suppose he had replied in this sense: Thinking is necessary in order to understand the empirically given, {\it and concepts and `categories' are necessary as indispensable elements of thinking.}'' \cite[p. 678]{Einstein65} (emphasis in the original)}
\end{quotation} 

Willingly or not, we physicists, are always producing our praxis {\it within} a specific representation. Representation is always first, experience and perception is necessarily second. Paraphrasing Wittgenstein's famous remark regarding language, the physical representation we inhabit presents the limits of the physical world we understand.\footnote{Let us notice, firstly, that ``physical'' should not be understood as a {\it given} ``material reality'', but rather as a procedure for representing reality in theoretical, both formal and conceptual, terms. And secondly, that the relation between such physical representation and reality is not something ``self evident''. The naive realist account according to which representation ``discovers'' an already ``fixed'' reality is not the only possibility that can be considered. A one-to-one correspondence relation between theory and reality is a very naive solution to the deep problem of relating theory and {\it physis.}} This marks a distance with respect to naive empiricism and positivism, also remarked by Einstein: 
\begin{quotation}
\noindent {\small ``I dislike [in a previous argumentation] the basic positivistic attitude, which from my point of view is untenable, and which seems to me to come to the same thing as Berkeley's principle, {\it esse est percipi.} `Being' is always something which is mentally constructed by us, that is, something which we freely posit (in the logical sense). The justification of such constructs does not lie in their derivation from what is given by the senses. Such a type of derivation (in the sense of logical deducibility) is nowhere to be had, not even in the domain of pre-scientific thinking. The justification of the constructs, which represent `reality' for us, lies alone in their quality of making intelligible what is sensorily given.'' \cite[p. 669]{Einstein65}}
\end{quotation} 

The naive empiricist viewpoint according to which there can be a ``direct access'' to the world that surround us by ``simply observing what is going on'' was fantastically addressed ---and ironically criticized--- by the Argentine Jorge Luis Borges in a beautiful short story called {\it Funes the memorious} \cite{Borges}. Borges recalls his encounter with Ireneo Funes, a young man from Fray Bentos who after having an accident become paralyzed. Since then Funes' perception and memory became infallible. According to Borges, the least important of his recollections was more minutely precise and more lively than our perception of a physical pleasure or a physical torment. However, as Borges also remarked: ``He was, let us not forget, almost incapable of general, platonic ideas. It was not only difficult for him to understand that the generic term dog embraced so many unlike specimens of differing sizes and different forms; he was disturbed by the fact that a dog at three-fourteen (seen in profile) should have the same name as the dog at three fifteen (seen from the front). [...] Without effort, he had learned English, French, Portuguese, Latin. I suspect, however, that he was not very capable of thought. To think is to forget differences, generalize, make abstractions. In the teeming world of Funes there were only details, almost immediate in their presence.''\footnote{The problem exposed by Borges is in fact, the same problem which positivists like Carnap, Nagel, Popper between many others tried to resolve without any success: the difficult relation between, on the one hand, phenomenological experience or observations, and on the other, concepts and theories. An interesting detailed and historical recognition of the many failures of the positivist project is \cite[Chap. 8]{Hempel65}.} Using the story as a {\it Gedankenexperiment} Borges shows why, for a radical empiricist like Funes, there is no reason why to assume a metaphysical identity between `the dog at three-fourteen (seen in profile)' and `the dog at three fifteen (seen from the front)'. For Funes ---the radical empiricist capable of apprehending experience beyond conceptual presuppositions--- there is no `dog', simply because experience does not contain the {\it moment of unity} required to make reference to {\it the same} through time. ``Locke, in the seventeenth century, postulated (and rejected) an impossible language in which each individual thing, each stone, each bird and each branch, would have its own name; Funes once projected an analogous language, but discarded it because it seemed too general to him, too ambiguous. In fact, Funes remembered not only every leaf of every tree of every wood, but also every one of the times he had perceived or imagined it.'' Existence, identity, non-contradiction, are ontological principles which provide the conceptual architectonic which allows us to connect the `the dog at three-fourteen (seen in profile)' and `the dog at three fifteen (seen from the front)' in terms of a {\it sameness}. It is only through these principles that we can think in terms of space-time systems ---such as, for example, a `dog'. Borges shows why  these principles are not self-evident {\it givens} of experience, and neither is a `dog'. And this is the reason why Borges also suspected that Funes ``was not very capable of thought.''

\smallskip

Going now back to quantum theory, the question we have investigated is the following: {\it Is it possible to provide an objective account of the orthodox quantum formalism?} Objective here should not be understood as ``a true reference to experience'' (whatever that means), but as the formal-conceptual preconditions under which we can represent a {\it moment of unity}, both in the mathematical level and the conceptual discursive level, within a physical theory. It is, in fact, this moment of unity which in turn allows for the famous {\it detachment} of the subject from the represented state of affairs. As we argued in \cite{deRondeMassri16}, in the case of the mathematical formalisms of physical theories the objective elements are given in relation to the notion of {\it invariance}. If we consider the invariant structure of the quantum formalism it is very easy to detect the ground for any objective consideration: it is obviously the Born rule \cite{deRonde16a, deRondeMassri18a}. What is not so easy of course, is to leave aside the deeply grounded (metaphysical) understanding of physical reality in terms of `systems' and `properties' inhabiting space-time and the actual realm of existence. According to our viewpoint, this actualist representation of physical reality ---when attempting to discuss and analyze what QM is really talking about--- has played the role of an {\it epistemological obstruction}  \cite{deRondeBontems11}. But if we stop trying systematically to relate QM to binary existence, to systems and properties, if we take seriously what the mathematical formalism is telling us, then we must accept ---following the footsteps of Wolfgang Pauli--- that the theory is pointing to the very reconsideration of the way in which we {\it represent} physical reality itself. Of course, in case we would be able to find such an objective representation for QM, the objective knowledge we posses cannot change after a known interaction. We agree with Schr\"odinger, that would be completely inadmissible for any physical theory. As we shall see in the following, the solution to this dilemma lyes inside the formalism of QM itself.

In \cite{Aerts84a}, Diederik Aerts discussed the EPR paradox as an {\it ad absurdum} proof making explicit ``the missing elements of reality in the description by quantum mechanics of separated physical systems.'' His deep and interesting analysis of the EPR paper can be nicely summarized in the following passage: 
\begin{quotation}
\noindent {\small ``[...] what E.P.R. show is the following

\smallskip

{\it Quantum mechanics describes correctly and in a complete way separated systems.

$\Rightarrow$ Quantities that are not compatible can have simultaneous reality.

$\Rightarrow$ Quantum mechanics is not complete.}

\smallskip

\noindent From this they can conclude that

\smallskip

{\bf D} {\it the quantum mechanical description of separated systems is not correct or not complete,} 

\noindent or

{\bf E} {\it quantities that are not compatible can have simultaneous reality and hence 
quantum mechanics is not complete.}

\smallskip

\noindent E.P.R. mention in the beginning of their paper that they suppose quantum 
mechanics to be correct. Hence they then also suppose quantum mechanics to 
give a correct description of separated systems. Two alternatives remain in this 
case: the quantum mechanical description of separated systems is incomplete or 
quantum mechanics is incomplete in the sense that quantities that are not 
compatible can have simultaneous reality. But in any case quantum mechanics is 
not complete. Hence E.P.R. can conclude that if quantum mechanics is correct, 
then it is not complete. What we showed in this paper is that the incompleteness 
arrives from {\bf D} and not from {\bf E}. Quantum mechanics is incomplete because it does 
not give a complete description of separated systems.''\cite[p. 427]{Aerts84a}}
\end{quotation}

That same year, Aerts proposed in \cite{Aerts84b} a solution to the paradox. Holding fast to the operational definition of an {\it element of physical reality}, this solution attempted to change QM ``in order to describe adequately separated systems''. The acceptance of the criteria of EPR physical reality goes in line with the operational quantum logic approach proposed by Constantin Piron \cite{Piron76} and subsequently developed by Aerts himself \cite{Aerts81}. In fact, Piron considered in detail this principle in order to define {\it actual properties} and {\it potential properties} within his own operational quantum logic \cite{Piron83}. As Aerts and Massimiliano Sassoli de Bianchi \cite[p. 20]{AertsSassoli17} ---both students of Piron--- explain with great clarity: ``the notion of `element of reality' is exactly what was meant by Einstein, Podolsky and Rosen, in their famous 1935 article. An element of reality is a state of prediction: a property of an entity that we know is actual, in the sense that, should we decide to observe it (i.e., to test its actuality), the outcome of the observation would be certainly successful.''  As we shall see in the following, the solution we propose to the EPR paradox within representational realism and the logos categorical approach involves a complete departure from the classical actualist (binary) representation of physics.

\section{Entangled Particles and Separable States}

%Given a PSA, $\Psi$, powers and potentia evolve deterministically, independently of actual effectuations, producing {\it potential effectuations} according to the following unitary transformation:
%\begin{equation}
%i \hbar \frac{d}{dt} \Psi (t) = H  \Psi (t),
%\end{equation}
%where $H$ is the Hamiltonian. While {\it potential effectuations} evolve according to the Schr\"odinger equation, {\it actual effectuations} are particular expressions of each power (that constitutes the PSA, $\Psi$) in the actual realm. The ratio of such expressions in actuality is determined by the potentia of each power.

Let us recall some definitions of the orthodox axiomatic account of QM. Let $\mathcal{H}$
be a separable Hilbert space. A density operator $\rho$  (i.e. a positive trace class 
operator with trace 1) is called a \emph{state}. Being positive (and self-adjoint), the eigenvalues of $\rho$ are non-negative and real and it is possible to diagonalize it. If the rank of $\rho$ is equal to 1, this diagonal matrix is given by $(1,0,\ldots,0)$ and $\rho$ is equal to  $vv^{\dag}$ for some normalized vector $v\in\mathcal{H}$. In this case, $\rho$ is called a \emph{pure state}. If the rank of $\rho$ is grater than 1 (or equivalently if $\mbox{Tr}(\rho^2)<1$), the state is called  a \emph{mixed state}. For example, the vector  $\alpha|0\rangle+\beta|1\rangle$, $|\alpha|^2+|\beta|^2=1$, gives the following density matrix:  
\[
\rho=\begin{pmatrix}
|\alpha|^2 & \alpha\overline{\beta}\\
\overline{\alpha}\beta&|\beta|^2
\end{pmatrix}.
\]
Notice that, if $\rho$ is a pure state (i.e., $\mbox{Tr}(\rho^2)=1$), there always exists a basis in which the matrix can be diagonalized as:
\[
\rho_{pure}=
\begin{pmatrix}
1 &0\\
0&0
\end{pmatrix}.
\]

The orthodox account of pure states is also supplemented by the following operational definition. If a quantum system is prepared in such way that one can devise a maximal test yielding with certainty a particular outcome, then it is said that the quantum system is in a \emph{pure state}. It is then stated that the pure state of a quantum system is described by a unit vector in a Hilbert space which in Dirac notation is denoted by $|\psi \rangle$.\footnote{As discussed in \cite{daCostadeRonde16, deRondeMassri18b} this definition is ambiguous due to the non-explicit reference to the basis in which the vector is written. It is this ambiguity which, in turn, mixes the notion of `state of a system' and `property of a system'.} This makes clear that it is only pure states which allow us to consider elements of physical reality in the EPR sense. Pure states guarantee the existence of an observable which is {\it certain} (probability equal to 1) if measured. Only pure states allow an interpretation of a quantum observable in terms of an {\it actual property}; i.e., a property that will yield the answer {\it yes} when being measured. On the very contrary, mixed states do not describe observables which, when measured, will be certain. When considering mixed states, all observables are {\it uncertain}; they all possess a probability which pertains to the open interval $(0,1)$. These properties are referred to in the literature as {\it indeterminate}  or {\it potential} properties. Indeterminate properties might, or might not become actualized in a future instant of time; they are {\it uncertain} properties which cannot be considered as elements of physical reality (in the EPR sense). As an example of a mixed state (i.e., $\mbox{Tr}(\rho^2)<1$) we can consider the following diagonal matrix,
\[
\rho_{mixed}=\begin{pmatrix}
\frac{1}{2} &0\\
0&\frac{1}{2}
\end{pmatrix}.
\]

\noindent Here, the observables related to the diagonal elements have probability $\frac{1}{2}$, which means that they will not be certain if measured.

\smallskip

Let us now consider states in $\mathcal{H} = \mathcal{H}_1\otimes \mathcal{H}_2$.  Pure states in  $\mathcal{H}_1\otimes \mathcal{H}_2$ can be given by any normalized vector, but in general, a state in $\mathcal{H}_1\otimes \mathcal{H}_2$ is given by a density operator in $B(\mathcal{H}_1\otimes \mathcal{H}_2)$. Among the states in $\mathcal{H}_1\otimes \mathcal{H}_2$ there is an orthodox distinction between: product states, entangled states and separable states. This distinction makes use of the idea that a Hilbert space represents a quantum system, and thus, that $\mathcal{H}_1\otimes \mathcal{H}_2$ represents two subsystems which only together give rise to the total system $\mathcal{H}$. However, the reductionistic logical structure implied by atomic systems is not captured by the orthodox Hilbert formalism. Regardless of this fact, the literature has nevertheless continued to think of the Hilbert mathematical structure in terms of atomic systems, making an implicit use of a reductionistic logic which, in fact, cannot apply to the formalism. 

A \emph{product state} is a state given by the tensor product of two states, $\rho=\rho_1\otimes\rho_2$, where $\rho_1$ is a state in $\mathcal{H}_1$ and $\rho_2$ is a state in $\mathcal{H}_2$.  The set of  \emph{separable states} in $\mathcal{H}_1\otimes \mathcal{H}_2$ is, by definition,  the closure (in trace norm) of  convex combinations of product states, $t_1\rho_{11}\otimes\rho_{21}+\ldots+t_n\rho_{1n}\otimes\rho_{2n}$ with $t_1+\ldots+t_n=1$ and $t_1,\ldots,t_n\in\mathbb{R}_{\ge 0}$. Any state outside the set of \emph{separable states} is called an \emph{entangled state}. For example, if $\alpha|0\rangle+\beta |1\rangle$ is a pure state in $\mathcal{H}_1$ and $\alpha'|0'\rangle+\beta'|1'\rangle$ is a pure state in $\mathcal{H}_2$, then the resulting product state in $\mathcal{H}_1\otimes \mathcal{H}_2$ is given by
\[
\rho_{pure}=\begin{pmatrix}
|\alpha|^2 & \alpha\overline{\beta}\\
\overline{\alpha}\beta&|\beta|^2
\end{pmatrix}\otimes
\begin{pmatrix}
|\alpha'|^2 & \alpha'\overline{\beta'}\\
\overline{\alpha'}\beta'&|\beta'|^2
\end{pmatrix}=
\begin{pmatrix}
|\alpha|^2|\alpha'|^2 & |\alpha|^2\alpha'\overline{\beta'}&\alpha\overline{\beta}|\alpha'|^2 & \alpha\overline{\beta}\alpha'\overline{\beta'}\\
|\alpha|^2\overline{\alpha'}\beta'&|\alpha|^2|\beta'|^2&\alpha\overline{\beta}\overline{\alpha'}\beta'&\alpha\overline{\beta}|\beta'|^2\\
\overline{\alpha}\beta|\alpha'|^2 & \overline{\alpha}\beta\alpha'\overline{\beta'}&|\beta|^2|\alpha'|^2 & |\beta|^2\alpha'\overline{\beta'}\\
\overline{\alpha}\beta\overline{\alpha'}\beta'&\overline{\alpha}\beta|\beta'|^2&
|\beta|^2\overline{\alpha'}\beta'&|\beta|^2|\beta'|^2
\end{pmatrix}.
\]

Given that we are working with pure states, the resulting product
state is pure and can be obtained with the vector 
$\alpha\alpha'|00'\rangle+
\alpha\beta'|01'\rangle+
\beta\alpha'|10'\rangle+
\beta\beta'|11'\rangle$.
In order to produce a separable state in 
$\mathcal{H}_1\otimes \mathcal{H}_2$, we can take a convex combination
of states as in the just mentioned example. For example,
\[
\rho_{sep}=\frac{1}{3}
\begin{pmatrix}
|\alpha|^2 & \alpha\overline{\beta}\\
\overline{\alpha}\beta&|\beta|^2
\end{pmatrix}\otimes
\begin{pmatrix}
|\alpha'|^2 & \alpha'\overline{\beta'}\\
\overline{\alpha'}\beta'&|\beta'|^2
\end{pmatrix}
+
\frac{2}{3}
\begin{pmatrix}
|a|^2 & a\overline{b}\\
\overline{a}b&|b|^2
\end{pmatrix}\otimes
\begin{pmatrix}
|a'|^2 & a'\overline{b'}\\
\overline{a'}b'&|b'|^2
\end{pmatrix}.
\]
On the contrary, a {\it maximally entangled state}, such as the Bell state will not be separable.  
\[
\rho_{ent}=\begin{pmatrix}
\frac{1}{2}&0&0&\frac{1}{2}\\
0&0&0&0\\
0&0&0&0\\
\frac{1}{2}&0&0&\frac{1}{2}
\end{pmatrix}.
\]

In the literature, the notion of {\it entanglement} has been defined in contradistinction to a notion of {\it separability}, which does not make reference to the space-time separability discussed by Einstein and has no clear physical counterpart. Furthermore, this mathematical feature ---namely, the closure of convex sums of product states--- plays no fundamental role within the quantum formalism. Its introduction only serves the purpose of supporting a fiction about the reference of QM to space-time separable ``elementary particles''. The notion of pure state also plays an essential role within the definition of separability, and consequently, also of entanglement.

\section{Logos Categorical QM: A Formal-Conceptual Approach}

The logos approach to QM presented in \cite{deRondeMassri18a, deRondeMassri18b} is able to explain in a visualizable manner through the use of graphs, how the objective character of the mathematical representation is restored when replacing the orthodox partial {\it binary valuations} by a {\it global intensive valuation}. By introducing this new type of existential quantification ---grounded on the invariance of the Born rule--- we were able, in turn, to derive a {\it Non-Contextuality Theorem} which shows how to escape Kochen-Secker contextuality and restore an objective reading of the mathematical formalism. But our approach is not only focused in the orthodox mathematical Hilbert scheme, it also stresses the need to supplement the formalism with {\it adequate physical concepts} which must be able to provide an {\it anschaulich} content and explanation of quantum phenomena.\footnote{The {\it anschaulich} aspect of physical theories was something repeatedly discussed by the founding fathers  of QM. More recently David Deutsch, taking distance from empiricists viewpoints which argue that theories are created from observations, has stressed the importance of their explanatory aspect \cite{Deutsch04, Deutsch16}.} Thus, by developing new (non-classical) notions we hope to explain in a new light the basic features already exposed by the orthodox quantum formalism. But before addressing the notion of {\it entanglement} more in detail from the perspective  of the logos approach to QM, let us first provide a general introduction to its basic elements.

\subsection{Beyond Actuality: Intensive Physical Reality and Objective Probability}  

As we have discussed in  \cite{deRondeMassri18a, deRondeMassri18b}, one of the main standpoints of the logos approach to QM is that by developing a notion of physical reality beyond binary existence ---imposed by the classical representation (in terms of `systems', `states' and `properties')---, it is in fact possible to provide a coherent objective representation of QM. This can be done, without changing the orthodox Hilbert formalism, without creating many unobservable worlds or introducing human consciousness within the analysis. According to this viewpoint, there is a main hypothesis presupposed within EPR's line of reasoning and argumentation which is wrong in a fundamental manner. QM simply does not describe an actual (separable) state of affairs. And it is the formalism itself which makes explicit this fact ---in many different ways--- right from the start: Heisenberg's indeterminacy principle, the superposition principle, Kochen-Specker theorem, Gleason's theorem, Born's probability rule, they are all ``road signs'' ---as Pauli used to call them--- that point in the direction of leaving behind the classical actualist representation of physics in order to understand the theory of quanta.

%In \cite{deRonde16a} it was argued that in order to take the formalism seriously, it is of outmost importance to reconsider quantum probability as providing objective knowledge of the state of affairs described by quantum theory. Here comes an important distinction with respect to orthodoxy which ---following positivism--- takes for granted the idea that physical theories are only mathematical devices capable of predicting the future observations of individual agents. On the very contrary, we have argued extensively that the realist path implies an account of theories as providing different {\it theoretical representations} ---both formal and conceptual--- of a state of affairs.\footnote{In particular, following Heisenberg's closed theory approach, we have argued that each physical theory provides a particular representation grounded on a mathematical formalism and a conceptual network \cite{deRonde16b}.} Each theory will provide a different characterization of a state of affairs. Each different representation will express the state of affairs in its own particular manner. The same state of affairs will be described in terms of particles by Newtonian mechanics, and in terms of charges and fields by Maxwell's electromagnetism. From this perspective, Born's rule implies the need of a deep reconsideration of the way in which physical reality is represented. It implies a necessary departure from the actualist binary (classical) representation of reality. 

It is argued today that physics can only describe `systems' with definite `states' and `properties'. This encapsulation of reality in terms of the classical paradigm ---mainly due to Bohr's philosophy of physics supplemented by 20th Century positivism--- has blocked the possibility to advance in the development of a new conceptual scheme. This is what David Deutsch \cite{Deutsch04} has rightly characterized as ``bad philosophy''; i.e., ``[a] philosophy that is not merely false, but actively prevents the growth of other knowledge.'' Taking distance from the Bohrian prohibition to consider physical reality beyond the theories of Newton and Maxwell, we have proposed the following extended definition of what can be naturally considered ---by simply taking into account the mathematical invariance of Hilbert formalism--- as a generalized element of (quantum) physical reality (see \cite{deRonde16a}).

\smallskip
\smallskip

\noindent {\it {\bf Generalized Element of Physical Reality:} If we can predict in any way (i.e., both probabilistically or with certainty) the value of a physical quantity, then there exists an element of reality corresponding to that quantity.}

\smallskip
\smallskip

\noindent As it will become clear, this redefinition implies a deep reconfiguration of the meaning of the quantum formalism and the type of predictions it provides. It also allows to understand Born's probabilistic rule in a new light; not as providing information about a (subjective) measurement result, but rather, as providing objective information of a theoretically described (potential) state of affairs. Objective probability does not mean that particles behave in an intrinsically random manner. Objective probability means that probability characterizes a feature of the conceptual representation accurately and independently of any subjective choice. This account of probability allows us to restore a representation in which the state of affairs is detached from the observer's choices to measure (or not) a particular property ---just like Einstein desired\footnote{As recalled by Pauli \cite[p. 122]{Pauli94}: ``{\it Einstein}'s opposition to [quantum mechanics] is again reflected in his papers which he published, at first in collaboration with {\it Rosen} and {\it Podolsky}, and later alone, as a critique of the concept of reality in quantum mechanics. We often discussed these questions together, and I invariably profited very greatly even when I could not agree with {\it Einstein}'s view. `Physics is after all the description of reality' he said to me, continuing, with a sarcastic glance in my direction `or should I perhaps say physics is the description of what one merely imagines?' This question clearly shows {\it Einstein}'s concern that the objective character of physics might be lost through a theory of the type of quantum mechanics, in that as a consequence of a wider conception of the objectivity of an explanation of nature the difference between physical reality and dream or hallucination might become blurred.''}. This means that within our account of QM ---contrary to the orthodox viewpoint---, the Born rule always provides complete knowledge of the state of affairs described quantum mechanically; in cases where the probability is equal to 1 and also in cases in which probability is different to 1. Or in other words, both {\it pure states} and {\it mixed states} provide {\it maximal knowledge} of a (quantum) state of affairs. Since there is no essential mathematical distinction, both type of states have to be equally considered; none of them is ``less real'', or ``less well defined'' than the other.

\subsection{The Logos Categorical Formalism}

Let us now recall  some basic mathematical notions of our logos categorical approach. We assume that the reader is familiar with the definition of a \emph{category}. Following \cite{deRondeMassri18a}, let $\mathcal{C}$ be a category and let $C$ be an object in $\mathcal{C}$. Let us define the category over $C$ denoted $\mathcal{C}|_C$.
Objects in $\mathcal{C}|_C$ are given by arrows to $C$, $p:X\rightarrow C$,  $q:Y\rightarrow C$, etc. Arrows $f:p\rightarrow q$
are commutative triangles,
\[
\xymatrix{
X\ar[rr]^f\ar[dr]_p& &Y\ar[dl]^q\\
&C
}
\]

\noindent For example, let $\mathcal{S}ets|_\mathbf{2}$ be the category of sets
over $\mathbf{2}$, where $\textbf{2}=\{0,1\}$ and $\mathcal{S}ets$ is
the category of sets.
Objects in $\mathcal{S}ets|_\mathbf{2}$
are functions from a set to $\{0,1\}$
and morphisms are commuting triangles, 
\[
\xymatrix{
X\ar[rr]^f\ar[dr]_p& &Y\ar[dl]^q\\
&\{0,1\}
}
\]
In the previous triangle, $p$ and $q$ are objects of 
$\mathcal{S}ets|_\mathbf{2}$
and $f$ is a function satisfying $qf=p$.

Our main interest is the category $\mathcal{G}ph|_{[0,1]}$ of graphs over the interval $[0,1]$. The category $\mathcal{G}ph|_{[0,1]}$ has very nice categorical properties \cite{quasitopoi, graphtheory}, and it is a \emph{logos}.  Let us begin by reviewing some properties of the category of graphs. A \emph{graph} is a set with a reflexive symmetric relation. The category of graphs extends naturally the category of sets and the category of aggregates (objects with an equivalence relation). A set is a graph without edges. An {\it aggregate} is a graph  in which the relation is transitive. More generally, we can assign to a category a graph, where the objects are the nodes of the graph and there is an edge between $A$ and $B$ if $\hom(A,B)\neq\emptyset$.
Given that in a category we have a composition law, the resulting graph is an aggregate.

\begin{definition}
We say that a graph $\mathcal{G}$ is \emph{complete} if there is an edge between two arbitrary nodes. A \emph{context} is a complete subgraph (or aggregate) inside $\mathcal{G}$. A \emph{maximal context} is a context not contained properly in another context. If we do 
not indicate the opposite, when we refer to contexts we will be implying maximal contexts.
\end{definition}

\noindent For example, let $P_1,P_2$ be two elements of a graph $\mathcal{G}$. 
Then, $\{P_1, P_2\}$ is a contexts if $P_1$ is related to $P_2$, $P_1\sim P_2$. Saying differently, if there exists an edge between $P_1$ and $P_2$. In general, a collection of elements $\{P_i\}_{i\in I}\subseteq \mathcal{G}$ determine a {\it context} if $P_i\sim P_j$ for all $i,j\in I$. Equivalently, if the subgraph with nodes $\{P_i\}_{i\in I}$ is complete. 

Given a Hilbert space $\mathcal{H}$, 
we can define naturally a graph $\mathcal{G}=\mathcal{G}(\mathcal{H})$
as follows. Following [{\it Op. cit.}] the nodes are interpreted as {\it immanent powers} and there exists an edge between  $P$ and $Q$ if and only if $[P,Q]=0$. This relation makes $\mathcal{G}$ a graph (the relation is not transitive). We call this relation {\it quantum commuting relation}.

\begin{theo}
Let $\mathcal{H}$ be a Hilbert space and let $\mathcal{G}$
be the graph of immanent powers with the commuting relation given by QM. 
It then follows that: 
\begin{enumerate}
\item The graph $\mathcal{G}$ contains all the contexts. 
\item Each context is capable of generating the whole graph $\mathcal{G}$.
\end{enumerate}
\end{theo}
\begin{proof}
See \cite{deRondeMassri18b}.
\qed
\end{proof}

\smallskip
\smallskip

As we mentioned earlier, an object in $\mathcal{G}ph|_{[0,1]}$ consists in  a map $\Psi:\mathcal{G}\rightarrow [0,1]$, where $\mathcal{G}$ is a graph. Then, in order to provide a map to the graph of immanent powers, we use the Born rule. To each \emph{power} $P\in\mathcal{G}$, we assign through the Born rule  the number $p=\Psi(P)$, where $p$ is a number between $0$ and $1$ called \emph{potentia}. As discussed in detail in \cite{deRondeMassri18a}, we call this  map $\Psi:\mathcal{G}\rightarrow [0,1]$ a {\it Potential State of Affairs} (PSA for short). Summarizing, we have the following:

\begin{definition}
Let $\mathcal{H}$ be Hilbert space and let $\rho$ be a density matrix.
Take $\mathcal{G}$ as the graph of immanent powers with the quantum commuting relation. 
To each immanent power $P\in\mathcal{G}$ apply the Born rule to get the number $\Psi(P)\in[0,1]$, which is called the potentia (or intensity) of the power $P$.  Then, $\Psi:\mathcal{G}\rightarrow [0,1]$
defines an object in $\mathcal{G}ph|_{[0,1]}$. We call this map a \emph{Potential State of Affairs}.
\end{definition}

Intuitively, we can picture a PSA as a table,
\[
\Psi:\mathcal{G}(\mathcal{H})\rightarrow[0,1],\quad
\Psi:
\left\{
\begin{array}{rcl}
P_1 &\rightarrow &p_1\\
P_2 &\rightarrow &p_2\\
P_3 &\rightarrow &p_3\\
  &\vdots&
\end{array}
\right.
\]

The introduction of {\it intensive valuations} allows us to derive a non-contextuality theorem that is able to escape Kochen-Specker contextuality \cite{deRondeMassri18a}). 

\begin{theo}
The knowledge of a PSA $\Psi$ is equivalent to the knowledge of the density matrix $\rho_{\Psi}$. In particular, if $\Psi$
is defined by a normalized vector $v_{\Psi}$, $\|v_{\Psi}\|=1$, then we can recover the vector from $\Psi$.
\end{theo}
\begin{proof}
See \cite{deRondeMassri18b}.
\qed
\end{proof}
\smallskip
\smallskip

Notice that our mathematical representation is objective in the sense that it relates, in a coherent manner and without internal contradictions, the multiple contexts (or aggregates) and the whole PSA. Contrary to the contextual (relativist) Bohrian ``complementarity solution'', there is in this case no need of a (subjective) choice of a particular context in order to define the ``physically real'' state of affairs. The state of affairs is described completely by the whole graph (or $\Psi$), and the contexts bear an invariant (objective) existence independently of any (subjective) choice. Let us remark that `objective' is not understood as a synonym of `real', but rather as providing the conditions of a theoretical representation in which all subjects are {\it detached} from the course of events. Contrary to Bohr's claim, in our account of QM, individual subjects are not considered as actors. Subjects are humble spectators and their choices do not change the objective representation provided by the theory.

\subsection{New Non-Classical Concepts}

Our approach seeks to understand quantum phenomena by introducing notions which are not only capable of matching the mathematical formalism in a natural manner, but also allow us to think about what is really going on according to the theory of quanta in a new light. As we have argued extensively \cite{deRonde16a, deRonde17a, deRonde17b, deRonde17c}, it is the concept of {\it immanent intensive power} which seems particularly well suited to describe what is going on according to the theory of quanta. Let us recall some important features of our new conceptual scheme. 

A immanent power, contrary to systems constituted by binary properties, has an {\it intensive existence}. A power is always quantified in terms of a {\it potentia} which measures its strength. The way to compute the potentia of each power is via the Born rule. Due to their invariant character, both powers and potentia can be regarded as being objective, meaning, independent of the (subjective) choice of any particular context. These notions allow us to escape the widespread story according to which, measurement in QM have a special status.\footnote{It is commonly argued that when we measure in QM, we always influence the quantum object under study ---which is just another way of making reference to the famous ``collapse'' of the quantum wave function. This idea, mainly due to Bohr's account of QM, implies that subjects define reality in an explicit manner; or as Bohr himself used to say: ``that in the great drama of [quantum] existence we are not only spectators but also actors.''} The Born rule is not anymore understood in epistemic terms, as making reference to the probability of obtaining a specific measurement outcome. Instead, Born's rule is now conceived as a way of computing an objective feature of the (potential) state of affairs represented by QM. The rule provides a definite value of the potentia of each power. The {\it immanent cause} allows us to argue that the single outcome found in a  measurement does not influence in any way the superposition as a whole: there is no collapse, no physical process taking place. Instead, the finding of an outcome is simply the path from an ontological description to an epistemic inquiry common to all physical theories (see \cite{deRonde17c}). The intensive account of powers also allows us to escape Kochen-Specker type contextuality which becomes in our scheme a purely epistemic feature, one that deals with the {\it epistemic incompatibility} of quantum experiments, and not with the {\it ontic incompatibility} of quantum existents (see \cite{deRonde16c, deRondeMassri18a}). 

As in any other physical theory, within the logos approach, the consideration of quantum measurements must be discussed independently of the mathematical formalism. The incomprehension of QM begun when the notion of measurement was explicitly introduced within the mathematical representation of the theory as an axiom (the projection postulate). It is this scrambling which has lead QM into all sorts of paradoxes and inconsistencies. Above all, it is responsible for having created the infamous measurement problem; a problem which has influenced in a decisive manner all the quantum mechanical research in the 20th Century.

\section{Quantum Entanglement in the Logos Approach}

Before discussing the logos' definition and understanding of {\it quantum entanglement}, let us sum up what we have learned up to now from our brief analysis of the history of the concept:
\begin{enumerate}
\item[I.] There is something really weird when thinking about the measurement of one of the two entangled quantum particles. If we follow the orthodox ---so called, ``minimal''--- interpretation of QM, the measurement of the one of the particles induces a ``collapse'' ---a real physical process--- not only of the measured entity, but also to the other distant partner. Something we could call a ``non-local super-luminous collapse'' or ---as Einstein named it--- a ``spooky action at a distance''.  
\item[II.] The correlation between space-time separated systems ---as part of the classical representation of physics--- is constrained by Boole-Bell type inequalities. 
\item[III.] The correlations appearing in Aspect's EPR quantum-type experiment violate the Boole-Bell inequality. Consequently, these correlations cannot be described by a classical (local-binary realistic) theory. This also means that the `clicks' cannot be represented from within a classical theory.
\item[IV.] QM reproduces theoretically the expected probabilistic predictions found within Aspect's EPR type-experiment. Something which ---let us remark--- is not related in any way to the violation of the inequality.
\item[V.] There is nothing, apart from a dogmatic metaphysical resistance to abandon the classical space-time atomist representation (inherited from Democritus metaphysics and Newtonian mechanics), indicating that QM is talking about ``small separable particles which inhabit space-time''. QM makes nowhere explicit use of this (metaphysical) conceptual representation. On the very contrary, everything in the formalism seems to point exactly in the opposite direction, namely, towards the conclusion that QM is not talking about ``particles in space time''.
\end{enumerate}

\noindent Given these facts, and the well known non-classical features of the orthodox quantum formalism, we ask the following questions: 
\begin{itemize}
\item  Why should we expect QM to be related to the classical representation inherited from space-time Newtonian mechanics and the metaphysics of (unobservable) atoms? 
\item Why should we keep using classical notions such as `system', `state' and `property', which we already know very well don't work at all in order to explain the theory of quanta? 
\item Why should we accept the naive positivist notion of observability ---and consequently, the collapse of the quantum wave function--- as a fundamental constraint in order to define physical reality? 
\end{itemize}

\noindent Independently of the answer to these questions, there are many hints coming from the analysis provided by the founding fathers which show a different path in order to develop QM beyond the classical (metaphysical) representation of physics. It is by taking seriously both the critical analysis put forward by Einstein and Schr\"odinger and the constructive conceptual approach suggested by Heisenberg's and Pauli's writings, that we have chosen to confront the Bohrian prohibition of developing new conceptual forms ---imposed by his doctrine of classical concepts \cite{Howard94b}--- and his reductionistic metaphysical presupposition according to which QM must be understood as a rational generalization of classical physics \cite{BokulichBokulich}. Up to now, {\it quantum entanglement}, following Bohr's restrictions of (classical) language and experience, has been systematically understood in terms of ``interacting elementary particles''. According to the logos approach, what we need to do ---above all--- is to think in a truly different manner. And this can be only done with the aid of a new conceptual scheme.

\subsection{Beyond Spookiness (... and Particle Metaphysics)}

In the history of science, many times, we physicists, have been confronted with ``spooky situations''. The ``spookiness'' always comes from the lack of understanding. Incomprehension of the unknown is always frightening. Just to give an example between many, the phenomena of electricity and magnetism were regarded as ``magical'' since the origin of humanity itself. Pieces of stone attracting each other without any material contact can be indeed ``spooky'', not to talk about lightnings coming from the sky. Until one day physicists were finally able to create a theory called electromagnetism which explained all these different phenomena. Physicists were even able to find out that these apparently different phenomena could be related in terms of a unified mathematical formalism. But it was only through the creation of the concept of `field', that we could finally grasp a deep qualitative understanding of these seemingly different phenomena. In the end, electricity and magnetism were two sides of the same represented physical reality. Suddenly, the spookiness had disappeared. 

Nobel laureate Steven Weinberg \cite{Weinberg}, when discussing about quantum entanglement, has argued that: ``What is surprising is that when you make a measurement of one particle you affect the state of the other particle, you change its state!'' This conclusion is indeed spooky, since we never observe in our macroscopic world that objects behave in such a strange manner. A table (or chair) in one place never affects the state of another distant table (or chair). If we do something to an object in a region of space $A$, there will be no action produced on an object situated in a distant region $B$. But we must also stress that Weinberg's amazement is implicitly grounded on two unjustified {\it ad hoc} assumptions, first, that QM talks about ``small particles'', and second, that quantum states suddenly ``collapse'' when being observed. 

The metaphysical picture provided by atomist Newtonian mechanics is of course a very heavy burden for today's quantum physics. Newton's metaphysical representation of the world has become the ``common sense'' of our time, advocated by many like a dogma that cannot be questioned. The worst part of this situation comes from those who do not even acknowledge that the Newtonian atomist picture {\it is} a metaphysical picture, and not an ``obvious'' or ``self evident'' way to talk about reality. But once we acknowledge that physics has always created new representations, an obvious question raises: could it be possible to create new concepts that would allow us to understand the phenomena implied by {\it quantum entanglement} in a manner which does not consider ``particles'', and which is not ``spooky''? Just in the same way as we created the notion of `field' in order to account for the phenomena of electricity and magnetism, could it be possible to develop a notion which is able to explain in an intuitive manner the quantum correlations that we encounter when performing measurements in the lab?

As we have discussed above, that which is responsible for the ``spookiness'' is the acceptance of two ---unjustified--- claims: first, the ``collapse'' of the quantum wave function; and second, the idea that QM talks about ``small particles''. Indeed, one does not find within the formal structure of QM a description or {\it reference} neither to `particles' nor to any strange `collapse process'. The notion of particle rather than helping understanding quantum phenomena seems to have played the role of an epistemological obstruction \cite{deRondeBontems11}. Also, as remarked by Dennis Dieks \cite[p. 120]{Dieks10}: ``Collapses constitute a process of evolution that conflicts with the evolution governed by the Schr\"{o}dinger equation. And this raises the question of exactly when during the measurement process such a collapse could take place or, in other words, of when the Schr\"{o}dinger equation is suspended. This question has become very urgent in the last couple of decades, during which sophisticated experiments have clearly demonstrated that in interaction processes on the sub-microscopic, microscopic and mesoscopic scales collapses are never encountered.'' In the last decades, the experimental research seems to confirm there is nothing like a ``real collapse'' suddenly happening when measurement takes place. Unfortunately, as Dieks \cite{Dieks18} also acknowledges: ``The evidence against collapses has not yet affected the textbook tradition, which has not questioned the status of collapses as a mechanism of evolution alongside unitary Schr\"odinger dynamics.'' 

At safe distance from many approaches which assume a classical metaphysical standpoint when analyzing QM ---introducing implicitly or explicitly classical notions within the theory---, the logos approach has been devised as an account of QM which stays close to the quantum formalism in the most strict manner. This implies for us, a suspicious attitude towards the (classical) notions of `system', `state' and `property'. Taking their place, we have created new (non-classical) concepts which attempt to satisfy the features of the quantum formalism ---and not the other way around, like for example in Bohm's approach to QM. According to the logos approach, QM talks about a potential realm which is independent of actuality. There is never a ``collapse'' of a quantum superposed state to a measurement outcome. QM talks about immanent powers with definite potentia, it does not talk about ``small particles''. From this standpoint, we have shown how through the aid of these notions we are able to explain the distance between the objective representation provided by the theory and the subjective measurements taking place {\it hic et nunc} in a lab  \cite{deRonde17c} ---dissolving in this way the infamous measurement riddle. The notion of {\it intensive power} provides an objective reference to the Born rule \cite{deRonde16a}, escaping the orthodox reading in terms of collapses and measurement outcomes. This new scheme has allowed us to provide an intuitive grasp of the meaning of quantum contextuality which abandons the idea that subjects ---as Bohr used to say--- are also actors in the great drama of (quantum) existence \cite{deRondeMassri18a}. The notion of quantum superposition also finds a natural {\it anschaulich} content within our proposal \cite{deRonde17a, deRondeMassri18b}. But can we also provide an intuitive grasp to the meaning of quantum entanglement?

According to our viewpoint, the main problem with the orthodox understanding of QM is that the representation of physical reality has been completely scrambled with a naive understanding of observability (for a detailed analysis see \cite{deRonde17a}). The infamous ``collapse'' scrambles the objective quantum theoretical representation with subjective epistemic observations. Because of this, the change in our (subjective) knowledge changes the theoretical description of (objective) reality itself.\footnote{This same scrambling takes place in the case of quantum contextuality and the so called ``basis problem''. See for a detailed analysis: \cite{deRonde16c}.} Hence, it is concluded that: ``measurement changes the state of affairs in an uncontrollable manner.'' But, as we have argued extensively, this is not necessarily the end of the story. If we accept that quantum probability is not making reference to measurement outcomes, but instead characterizes an objective feature of the state of affairs represented by QM, then there is no need of considering epistemic measurements at all within the theoretical representation. In such case, it is possible to restore the consideration of quantum phenomena ---without the intromission of epistemic knowledge--- as part of an objective representation of quantum physical reality.

\subsection{Quantum Entanglement Beyond Purity and Separability} 

Classically, the coding of a message implies no action at a distance. If we take two envelops and put in one of them a `red' piece of paper, and a `blue' piece of paper in the other one; and we then share the two envelops between two partners which travel to distant places like Buenos Aires and Cagliari; the moment one of the partners in Buenos Aires opens the envelop, he will learn not only which was the paper in his own envelop, he will also learn simultaneously what is the color of the paper within the envelop of his partner in Italy. There is of course nothing spooky about this. None of the pieces of paper traveled from Buenos Aires to Caglairi or vice versa, just because one of the two partners opened the envelop. The state of affairs didn't change. What changed, when opening the envelope, was the knowledge of the state of affairs represented in classical terms. As we shall see, according to the logos approach, the main difference with quantum entanglement is the type of coding: while classical entanglement codifies {\it actual binary information}, QM is able to codify {\it potential intensive information}. Let's discuss in detail what we mean by this. 

First rule of the logos approach: forget about ``small particles''. Forget about `systems' constituted by `properties' and `states'. And  for the sake of the argument, let us accept that QM talks about something else, namely, about immanent intensive powers (see \cite{deRonde16a, deRonde17a, deRonde17b, deRonde17c}). According to this explanation, quantum entanglement is just a way to codify potential intensive powers (instead of actual pieces of paper). QM provides a new way, different to the classical one, to codify information  grounded on a {\it relational-coding} of information within the potential realm. As it should be clear from section 6.1, the potential-coding is already present when different aggregates of powers are considered within the same quantum situation. A compositional situation is not separable in the intuitive sense of classical systems discussed by Einstein. The choice of this term, `separable', is extremely inappropriate since the closure (in trace norm) of convex combinations of product states is a mathematical feature which does not have an obvious physical counterpart. In QM, there is no reductionism when considering the whole and its parts already in this most simple compositional situation, and thus, the application of the notion of separability becomes completely untenable. A compositional situation is not separable in the classical space-time sense of the term. Separability in classical physics is always understood in reductionistic terms. Two particles coming together will be described jointly, by just adding the particles. But this is not the case in QM where the addition of powers produces interactions which are not just the sum of each set of powers. The distinction between {\it pure states} and {\it mixed states} is also completely superfluous from a purely mathematical perspective.\footnote{This point has been already addressed by David Mermin in \cite[p. 758]{Mermin98}.} This distinction has been introduced by another metaphysical dogma which considers that only {\it certain} (measurement) predictions can be related to real (actual) physical quantities ---through EPR's criteria of {\it element of physical reality}. 

%In the logos approach, since all matrices describe in a correct manner an aggregate of powers and potentia, we will never make any use of such a distinction. 

%If we visualize a {\it pure state} and a {\it mixed state} using graphs we obtain the following: 
%\begin{center}
%\includegraphics[width=12em]{pure.png}\quad
%\includegraphics[width=12em]{mixed.png}\quad
%\end{center}

%\noindent If we produce the graphs of a separable state and an entangled state we obtain:
%\begin{center}
%\includegraphics[width=12em]{separable.png}\quad
%\includegraphics[width=12em]{entangled.png}
%\end{center}

%\noindent We see there is no essential difference between these graphs. All of them are just aggregates of nodes (interpreted as powers) with an intensive valuation (see \cite{deRondeMassri18a}). 

These two distinctions, pure-mixed and separable-entangled ---widespread in the specialized literature---, are simply irrelevant from a purely mathematical perspective. Both distinctions have been introduced within the theory attempting to sustain a metaphysically unjustified story according to which QM talks about space-time {\it separable} atomic systems described in terms of definite valued properties giving rise to {\it certain} predictions (probability = 1). While pure states are strictly related to the possibility of producing measurement outcomes with {\it certainty}, separable states are intuitively grounded on the presupposition that QM talks about {\it space-time separable} systems. However, none of these conceptual requirements is compatible with the orthodox quantum formalism. In fact, if one looks at the formalism and considers two PSAs, the composition of them produces the product of powers ---not the sum, as in the reductionistic case of systems. This can be understood in a very intuitive manner. If we have two apparatuses in a room, each one of them must be understood according to the logos as an aggregate of powers, say $N$. If we consider them jointly, meaning, if we use them together in order to perform new experiments, the number of powers increase quadratically, as $N^2$. The new possibilities are not the sum of the previous ones, they are much more. Obviously, adding two apparatuses allows many more possibilities than just the sum of their previous possibilities. But this is clearly not the case when considering atoms and their properties. If we add two systems, the properties are summed instead of multiplied. Given two systems with a number of properties, $R$ and $R'$, respectively; their joint consideration is just the sum of the properties of each system, namely, $R+R'$.\footnote{A paradigmatic example is the completely inelastic crash of two systems. While before the crash the two particles are separated and their mass are $m$ and $m'$, and their velocities are $v$ and $v'$, respectively; after the crash they become a (non-separable) single system of mass $m+m'$ with a common velocity $v_f$. The essential property characterizing the two systems ---namely, their mass--- becomes nothing else than the sum of masses.} But, as we know, there is an essential difference when considering the addition of sets in classical mechanics and the addition of vector spaces in QM. If we take two rays which intersect each other, in terms of classical set theory, the addition of the rays is just the two rays; however, in terms of vector spaces the addition of two rays (now considered as subspaces) is more than just their sum, it is the whole plane generated by the two rays (see also \cite{Aerts81}).\footnote{While orthodoxy interprets Hilbert spaces following atomist metaphysics as `physical systems' (i.e. as elementary particles), the logos approach understands configuration space in relation to the possible actions that can be performed.} 

The logos approach imposes the need to redefine the notions of entanglement and separability going necessarily beyond the distinction between pure-mixed states, and the notion of space-time separability.

\subsection{Relational Potential Coding (It Might be Spooky, But There is No Action)} 

As discussed in the EPR paper, two powers can be related in terms of a {\it definite value}. In his papers, Schr\"odinger also remarks that when considering $X = x_2 - x_1$ or $P = p_2 - p_1$,  $X$ and $P$ might posses a definite value, say $X'$ and $P'$, but their relata, $x_1, x_2, p_1$ and $p_2$ do not. 

\begin{quotation}
\noindent {\small ``[...] the result of measuring $p_1$ serves to predict the result for $p_2$ and vice versa [since P' is known]. But of course every one of the four observations in question, when actually performed, disentangles the systems, furnishing each of them with an independent representative of its own. A second observation, whatever it is and on whichever system it is executed, produces no further change in the representative of the other system.

Yet since I can predict {\it either} $x_1$ {\it or} $p_1$ without interfering with system No. 1 and
since system No. 1, like a scholar in examination, cannot possibly know which of the two questions I am going to ask it first: it so seems that our scholar is prepared to give the right answer to the first question he is asked, anyhow. Therefore he must know both answers; which is an amazing knowledge, quite irrespective of the fact that after having given his first answer our scholar is invariably so disconcerted or tired out, that all the following answers are "wrong'' \cite[p. 555]{Schr35b}}
\end{quotation}

\noindent This relational aspect of QM is not shared by (reductionistic) classical theories grounded on set theory and an their underlying Boolean logic. What is difficult to accept from a classical metaphysical viewpoint is that a relation can have a value without,  at the same time, their relata possessing one. This first level of relationalism applies to actual observations. The result of the outcome $x_1$ is correlated (or anti-correlated) to the actual outcome $x_2$ (and, of course, viceversa). All this is very well known. However, this is not the only type of relation implied by QM. Quantum relationalism ---which will be addressed in detail in \cite{deRondeFMMassri18a}--- allows a potential coding of powers in two different levels. While in the first level of representation we have {\it definite} or {\it effective relations} dealing with actual measurements, in the second ---most important--- level we have an {\it intensive relation} between powers which requires, at the epistemic level, a statistical type of analysis. It is this level, which we consider to be the most characteristic of QM. These relations have been overlooked by the physics community which has embraced the positivist-empiricist obsession towards measurement outcomes and the existence of small unobservable particles.

Due to its restricted focus on measurement outcomes, it is only effective relations which have been considered and analyzed by the community discussing quantum information processing. Let us provide a definition of such relations:

\medskip

\noindent {\it
{\sc Effective Relations:} The relations determined by a difference of possible actual effectuations. Effective relations discuss the possibility of an actualist definite coding. It involves the path from intensive relations to definite correlated (or anti-correlated) outcomes. They are determined by a binary valuation of the whole graph in which only one node is considered as true, while the rest are considered as false.}

\medskip

\noindent The potential coding making reference to the potentia of correlated powers must be analyzed in a statistical manner. What needs to be studied in detail is the way in which the potentia of such correlated powers is able to interact in the potential realm. Intensive relations are, according to the logos approach, the true access-key to quantum information processing. 

\medskip

\noindent {\it
{\sc Intensive Relations:} The relations determined by the intensity of different powers. Intensive relations imply the possibility of a potential intensive coding. They are determined by the correlation of intensive valuations.}

\medskip

\noindent In the logos approach it is possible to consider, within a single graph the entanglement of powers and both intensive and effective relations. The quantum situation $\Psi_1|_{C_1}$ exposes on the one hand the statistical correlation of powers 0 and 1, and on the other, the fact that every time we measure, we will obtain correlated outcomes, 0 and 0 or 1 and 1. This means that while intensive relations relate to intensive valuations, effective relations relate to effective valuations.  

\smallskip

Before we define mathematically the concept of \emph{effective valuation}, 
let us recall the definitions of intensive and 
binary valuations.
An \emph{intensive valuation} (or a PSA) is a map $\Psi:\mathcal{G}\to[0,1]$ 
and a \emph{binary valuation} is a local map $\nu:\mathcal{G}\to\{0,1\}$
both compatible with the structure of the graph. 
Recall that global binary valuations do not exist, 
but global intensive valuations do, see \cite{deRondeMassri18a}.

\begin{definition}
Let $\Psi:\mathcal{G}\to[0,1]$ be a PSA
and let $\mathcal{C}\in \mathcal{G}$ be a context.
An \emph{effective valuation} over $\mathcal{C}$ is a random variable $\nu$
with range $\{0,1\}$ which takes $P_k = 1$ and the rest $P_i=0$ (for all $i \neq k$).
\end{definition}
Notice that the concept of effective valuation depends on a context.
Now that we have the mathematical definition of effective valuation, 
let us give the mathematical definition of intensive and 
effective relations.

\begin{definition}
Let $\Psi_1$ and $\Psi_2$ be two PSA.
We say that $\Psi_1$ and $\Psi_2$ are \emph{related intensively}
if there exists an isomorphism $\tau$ making the following diagram commute
\[
\xymatrix{
\mathcal{G}_1\ar[rr]^{\tau}\ar[dr]_{\Psi_1}& &\mathcal{G}_2\ar[dl]^{\Psi_2}\\
&[0,1]
}
\]
\end{definition}
For example, in the next picture we can visualize the intensive relation
between two PSA,
\begin{center}
\includegraphics[width=8em]{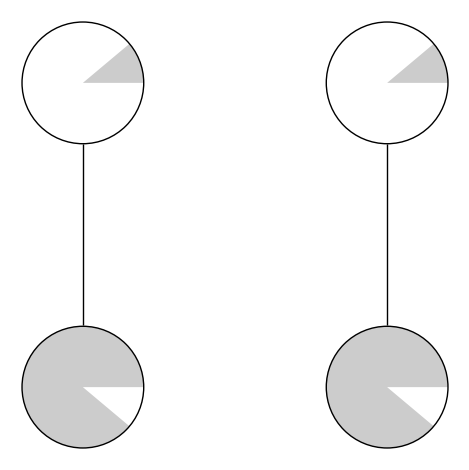}
\end{center}
Notice that an effective valuation applied to the first graph
is different from an effective valuation applied to the second graph.

\begin{definition}
Let $\Psi_1$ and $\Psi_2$ be two PSA.
We say that $\Psi_1$ and $\Psi_2$ are \emph{related effectively}
if every effective valuation on $\mathcal{G}_1$ is
correlated (or anti-correlated) to an effective valuation 
on $\mathcal{G}_2$. If $\Psi_1$ is 
related effectively to $\Psi_2$, we picture a two-way arrow
between the two graphs.
\end{definition}
For example, the following picture denotes an effective relation,
\begin{center}
\includegraphics[width=8em]{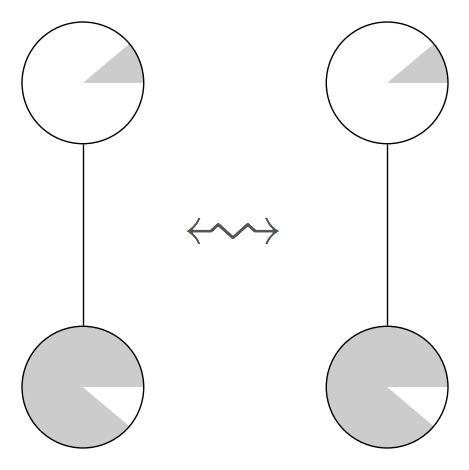}
\end{center}

\noindent Let us also remark that if there exists an effective relation between two PSAs, then, consequently, there will also exist an intensive relation between them. But the inverse does not follow. 
 
{\small $$  EFFECTIVE \ RELATIONS \ \ \Rightarrow \ \ INTENSIVE  \ RELATIONS$$ 
$$ INTENSIVE \ RELATIONS \ \ \nRightarrow \ \ EFFECTIVE \ RELATIONS$$} 

\noindent Through the notions of intensive relation and effective relation we are now able to provide a new definition of entanglement and separability. A definition which does not depend on the distinction between pure and mixed states nor makes any reference whatsoever to space-time separability. 

\begin{definition}
Let us introduce the following definitions
for $\Psi_1$ and $\Psi_2$ two PSAs, 
\begin{itemize}
\item \emph{Strong Quantum Entanglement}: If $\Psi_1$ and $\Psi_2$ are related effectively
and intensively.
\item \emph{Weak Quantum Entanglement}: If $\Psi_1$ and $\Psi_2$ are related intensively
but not effectively.
\item \emph{Separable}: If $\Psi_1$ and $\Psi_2$ are not related effectively
nor intensively.
\end{itemize}
For example, in the previous two pictures, the first one depicts a weak entanglement between two
PSAs and the second a strong entanglement.
\end{definition}

%
%Two labs which are not part of the same PSA can be considered as separable. 
%This means that even though they might share the same intensive relations, 
%they will not have the same effective valuations. 

The ontological account of QM in terms of intensive powers makes explicit that the most important coding of quantum information lies in the intensive statistical level of analysis, not in the level of single measurement outcomes. What needs to be codified is the relational potentia of powers, not the specific outcomes. Adding to intensive relations, effective relations show that quantum relations are not merely ``statistical'' in the classical sense, they are ---as explicitly tested in Aspect's experiments--- much stronger. Unfortunately, restricted by the almost exclusive reference to measurement outcomes and atomist metaphysics, the field has limited itself to the analysis of the actual measurement effectuations of ``two entangled space-time separated particles''. This is not difficult to understand given the widespread ---either implicit or explicit--- belief in ``collapses'' and the strict reference to the actual realm of existence ---either as referring to particles or to measurement outcomes. What we need to do now, according to the analysis provided by the logos approach, is to advance towards the understanding of the codification of potentia. This must be done not only in formal categorical terms but also in conceptual representational terms. Only together, the mathematical formalism and physical concepts can provide a truly {\it anschaulich} content to QM.

\section{The {\it Anschaulich} Content of Quantum Entanglement} 

From the logos standpoint, we need to answer the following question: What is quantum entanglement in conceptual terms? Or in other words, how can we think intuitively about this physical feature mathematically represented by the quantum formalism in terms of vectors and tensor products? Any adequate account of QM should be able to answer these questions in an intuitive manner. In order to do so, we must abandon the orthodox fictional story of interconnected particles collapsing their states in a super-luminous manner. This story does not make any sense. An adequate conceptual representation should be able to provide an {\it anschaulich} access and the possibility of thinking about the phenomena described by the theory. We believe that the logos approach is able, thorough the addition of new concepts, to produce such a conceptual insight to quantum entanglement. Let us discuss this in some detail. 

In order to grasp the meaning of entanglement we will use the examples already provided by two different sports: football ---as we all know it--- and United States' football. Let us begin by football. Argentine Lionel Messi, Brazilian Neymar da Silva Santos J\'unior and Uruguayan Luis Suarez played together in season 2016 of the Spanish Football Ligue at an amazing level. But what was taking place for this to happen? Apart from the fact that Messi, Neymar and Suarez are three of the best players in the world, there are very good players which simply don't play well together. This phenomena is well known, the addition of great players doesn't necessarily create a  great team. Indeed, in football the creation of a team is a difficult process which is not at all understandable through a purely reductionistic type of logic. It is not the case that if you have 11 players with the capabilities of Messi, or Diego Armando Maradona, you will necessarily end up with an amazing team. Unlike other sports, the powers of players within a football team are highly contextual and relational. Each position within the field has a very specific need, a particular demand. The powers of a player in one position, say, in the defense, are different from those required in the attack. The aggregate of powers of a player in the left hand side of the field (like Neymar) are different from those playing in the right hand side of the field (like Messi) or even in the middle (like Suarez). Messi is the best player in the world, but in {\it his} specific position ---which, of course, turns out to be one of the most important positions within football due to its proximity to the goal. However, we suspect that Messi would not be such an incredible player acting as a goalkeeper. Messi simply doesn't seem to have the specific powers required in order to act defending the goal in an adequate manner. As we have discussed elsewhere, immanent powers are contextual existents. It is due to this contextual feature of powers that a football team ---as all football fans know--- is something truly difficult to conform. Producing a team implies a process of creating a balanced and capable `global individual', namely, the team itself. 

An interesting point for our analysis of quantum entanglement is the way in which the correlations of teams are built up. The process through which a team learns how to play together; or in other words, how to ``play as a team''. This means nothing else than to conform actions in a completely correlated manner, as a unity, as an individual. The answer to this question is obvious for anyone how has played a sport: you get more correlated by simply playing together, by training, by interacting through long periods of time. The more a team practices together, the more it will be able to correlate in an adequate manner the relations between the different aggregates of powers the team is composed of. The more a team trains together, the more they interact, the better they will correlate the actual effectuations of powers the day of the game. The interaction between the players creates something which we could also call ``entanglement''; i.e., a potential coding of possible moves of players within the football field. 

The first type of relation is a {\it definite potential coding} which involves the preparation of very specific moves within determined situations. These are called in Argentine ``jugadas preparadas'' ---which translated means ``prepared moves''. Moves prepared in order to deal with corner kicks or free kicks near scoring positions. But there is a much more interesting type of potential correlation to which football fans ---at least in Argentina--- call ``juegan de memoria'', which means that players are able ``to play by heart''. We will call this, {\it intensive potential coding}. Intuitively, it means that some teams are so accustomed in playing together that they act in a completely correlated manner, and this is a kind of non-local behavior. The entanglement of the powers of Messi, Suarez and Neymar gets more correlated the more the players interrelate, the more they practice together. It is only then, that they can acquire an entangled relationship in which they are able to produce interrelated actions without a previous written plan. For example, when Messi goes to the left, Suarez already knows that he has to go to the right. When Neymar attacks by the left, Messi knows he has to go to the center. The performances that we, football fans, witness every Sunday are highly correlated actions which are not written beforehand. Most of these actions, unlike the case of U.S. football, are not previously determined. The intensive potential coding implies a truly potential correlated aggregate of actions which are not written anywhere. And this is why there is a lot of ``instant creation'' within the football field. And also, why football is so interesting. It is also interesting to notice that the analysis of potential coding does not require effective actualizations. We know that the national team of Argentina and Brazil are better football teams that those of Australia or New Zeland. But this does not mean that Argentina and Brazil will always beat Australia and New Zeland. 

U.S. football and baseball are much more strict regarding the possibilities of creation. The space for the unknown action is much more restricted by the structure of the game. In U.S. football the {\it definite potential coding} of possible moves has a much more important role than in football. In this case the strategy in each situation becomes of outmost importance. This is because U.S. football is a discrete set of situations. In each situation, the trainer must choose only one between a set of already prepared possible actions. An action is performed and then a new action is required. There is no continuity in the game and the trainer has to become a strategic leader. In this case, the U.S. football player is more a soldier following orders than an artist creating a movement. 

%In U.S. football a game can be viewed as a measure of the entanglement of a team's. The combined actions that need to be performed within the field require the previous creation of interrelated powers, the more interaction, the more correlations can be controlled. The trainer of the team requires a determinate codification of potential powers which allows for controlled actual processes.   

When we go to see a football match, we see how the entanglement encapsulated in potential correlations is actualized during the game. There is nothing spooky about this. As there should be nothing spooky about QM. The potential realm comprises the relations between intensive powers. Against the expectations of Einstein, what is going on cannot be thought within the classical space-time representation. Potential reality cannot be encapsulated in terms of classical notions. Immanent intensive powers are simply not space-time existents. They are non-separable since interaction and entanglement is always present. The analysis of locality is very restricted and has to be discussed recalling this main aspect of QM. 

We can now understand what is the main problem of EPR's hypothesis, namely, the presupposition that QM talks about separable quantum particles that inhabit space-time. The use of the notion of `particle' within quantum entanglement does not create understanding but rather the opposite, it creates spookiness.  

%TENSCION BETWEEN STATISTICAL RELATIONS AND ACTUAL EFFECTUATIONS. 

\section{Restoring an Objective Account of Entanglement}

In the orthodox von Neumann-Dirac axiomatic formulation of QM, the mathematical formalism is scrambled with subjective measurement observations right from the start. This is done through the explicit introduction of the projection postulate which implies the ``collapse'' of the quantum wave function, and in turn, is also responsible for the creation of the infamous measurement problem. This pseudo-problem has hunted QM since the early debates of the founding fathers. 

By reconsidering the Born rule as an objective feature of the formal representation itself (of a potential state of affairs), the logos approach is detached from the the idea of a ``collapse'' right from the start. Quantum probability is not making reference to single measurement outcomes, it provides a way to compute a specific feature characterizing the potential state of affairs; namely, the definite potentia of each power. In this respect, in the logos approach, contrary to the orthodox account, the state `$|0 \rangle$' and the state `$\frac{1}{\sqrt{2}} (| 0 \rangle + |1 \rangle)$' are both providing exactly the same complete knowledge of the (potential) state of affairs described by quantum theory. Contrary to the orthodox account ---grounded on EPR's {\it element of physical reality}---, these two states provide the same type of definite knowledge.\footnote{This is the main sin of positivism giving rise to the measurement problem and a century of confusions and misunderstandings regarding the meaning of quantum theory.} Both states provide an equally objective representation of a well defined quantum state of affairs. This is due to the fact ---let us remark once again--- that the number accompanying each power makes reference to a potentia, and not to a measurement outcome. We could say that, contrary to the classical case, the state `$|0 \rangle$' is not ``more certain'' than the state `$\frac{1}{\sqrt{2}} (| 0 \rangle + |1 \rangle)$'; both states are as certain as they can be, they both provide complete information. We could talk here of an ontological shift from `binary (actualist) certainty' to an `intensive (potential) certainty'. 

Within the orthodox understanding of {\it entanglement}, the collapse of the quantum state has played a fundamental role within the description of the state of affairs. The collapse is the main responsible for scrambling together the objective representation provided by the theory with the epistemic acquisition of knowledge provided via measurements in the lab. Within the logos approach, we have unscrambled the objective representation provided by the mathematical formalism from the subjective epistemic process of measurement right from the start. We have done this by taking seriously the formalism of QM. This means for us to pay attention only to what the mathematical formalism is telling us regardless of metaphysical (atomist and actualist) presuppositions. 

Any mathematical formalism that attempts to become a physical theory which is able to represent an objective state of affairs must look to the invariant structure of the mathematical formalism \cite{deRondeMassri16}. It is there where you can find the elements that can be understood independently of the reference frame, and consequently, of empirical subjects observing the represented state of affairs. One of the main differences between QM and classical physical theories is that while in the latter case invariant quantities are binary, in the former invariant quantities ---computed via the Born rule--- relate to a number within the interval $[0,1]$. As we have argued in \cite{deRonde16a}, this mathematical feature has a radical consequence in the way physical reality is represented. The shift we are making reference to is that going from a {\it binary representation} into an {\it intensive representation} of physical reality. If we bite the bullet and accept this radical shift regarding our understanding of reality ---from an actual representation to a potential representation---, it is then possible to produce an objective account of what QM is really talking about. An explicit proof of this fact is the {\it Intensive Non-Contextuality Theorem} derived in \cite{deRondeMassri18a} which allows us to escape Kochen-Specker contextuality. As argued in \cite{deRonde17c}, by reconsidering physical reality in terms of a potential realm independent of actuality we can also escape the measurement problem. 

In the logos approach, measurement outcomes (or observations) are not considered as part of the theory; epistemic results are not themselves part of theoretical representations \cite{deRonde16a, deRonde17a, deRondeMassri18b}. Any trained physicist knows there is a huge gap between the work done by theoretical physicists and experimentalists. What we need to stress at this point is that the epistemic process of measurement is not contained within the mathematical formalism of any theory. Physical theories do not teach us how to measure. Theories only provide us with a formal-conceptual representation of a state of affairs, period. While classical mechanics provides a complete representation in terms of `interacting particles', electromagnetism describes situations  in terms of `interacting electro-magnetic waves'. Any physicist knows very well that none of these theories contain an explanation of how to measure a `particle' or a `field'. Theories simply do not come with a user's manual explaining the technical subtleties of how to measure things in the lab.   

We agree with Schr\"odinger and Einstein: objective knowledge of a state of affairs must rest exactly the same before and after a known interaction. The {\it theoretical knowledge} of the objective representation of a state of affairs should not be confused with the {\it epistemic knowledge} acquired within the lab. Any interaction must be considered in theoretical terms. In this respect, the logos approach goes against the Bohrian orthodox claim according to which our knowledge (or observation) of a quantum system changes physical reality itself. The logos approach restores Einstein's requirement of a detached observer for QM. It confronts in this way the Borian idea according to which ``we are not only spectators but also actors in the great drama of existence''.

%\noindent {\it {\sc Objective Theoretical Knowledge:} The knowledge provided by an objective physical representation. Interaction is the way in which the elements described by the representation relate between each other. For example, a crash of particles as described by classical mechanics or the interaction of two fields according to electromagnetism. The knowledge provided by the representation is always objective ---in the sense that the subject is completely {\it detached} from the represented state of affairs--- and complete ---in the sense that it involves no epistemic procedure.}

%\noindent {\it  {\sc Subjective Empirical Knowledge:} The knowledge of empirical subjects (or agents) collected through measurements within a lab by presupposing a specific theoretical representation of the state of affairs. The epistemic praxis of measuring a physical quantity is not contained in the mathematical formalism. Epistemic knowledge can be complete or incomplete depending of the information collected by the agent.}

Our claim is that QM provides a complete representation of a potential state of affairs in terms of immanent powers with definite potentia. Since we are dealing with an intensive (non-binary) realm of reality, in order to learn about a specific quantum state of affairs present in the lab we always need to perform a repeated series of (statistical) measurements. A single outcome simply does not provide enough information in order to understand what is going on according to the theory of quanta. In this way, the logos approach is able to introduce a new non-classical account of physical reality which restores the distance between the knowledge provided by an objective theoretical representation and the subjective epistemic knowledge produced by measurements.

\section*{Conclusion}

In this paper we have critically argued that the distinctions between pure-mixed and entangled-separable, derived from classical metaphysical standpoints ---such as binary existence and the reference to small particles--- are not only superfluous and irrelevant when considering the orthodox mathematical formalism of QM, but also problematic. Leaving aside these distinctions, we have reconsidered the notions of {\it quantum entanglement} and separability from the perspective of the logos categorical approach \cite{deRondeMassri18a, deRondeMassri18b} providing a new definition of entanglement in terms of intensive and effective relations. Entanglement is defined in completely objective terms without considering measurements or collapses. These new definitions make no use whatsoever of the distinction between pure ad mixed states, nor any reference to space-time separability. Finally, we have also discussed the {\it anschaulich} content of quantum entanglement providing a new understanding which escapes its ``spookynes'' (or action at a distance).

\section*{Acknowledgements} 

C. de Ronde would like to thank Don Howard for historical references. We would also like to thank Dirk Aerts, Massimiliano Sassoli de Bianchi, Giuseppe Sergioli, Hector Freytes and Raimundo Fernandez Moujan for discussions on related subjects. This work was partially supported by the following grants: FWO project G.0405.08 and FWO-research community W0.030.06. CONICET RES. 4541-12 and the Project PIO-CONICET-UNAJ (15520150100008CO) ``Quantum Superpositions in Quantum Information Processing''.

\end{document}